\documentclass{aa}
\usepackage{graphicx}
\usepackage{txfonts}

\newcommand{\Teff}{T_{\rm eff}}

\newcommand{\Mn}[5]{\mbox{$#1\,^#2{\rm #3}^{{\rm #4}}_{\rm #5}$}}
\newcommand{\mA}{{\rm m\AA}}
\newcommand{\Elow}{E_{\rm low}}

\newcommand{\Vmic}{\xi_{\rm t}}

\newcommand{\loggfe}{\log (gf\varepsilon)}
\newcommand{\loggferef}{\log (gf\varepsilon)_{\rm ref}}
\newcommand{\loggfestar}{\log (gf\varepsilon)_{\ast}}
\newcommand{\loggfesun}{\log (gf\varepsilon)_{\odot}}
\newcommand{\loge}{\log\varepsilon}
\newcommand{\kms}{km s$^{-1}$}
\newcommand{\SH}{S\!_{\rm H}}           

\begin{document}
\title{NLTE abundances of Mn in a sample of metal-poor stars
\thanks{Based on observations collected at the European Southern Observatory,
Chile, 67.D-0086A, and the Calar Alto Observatory, Spain.}}
\author{M. Bergemann\thanks{Research
supported by the International Max Planck Research School (IMPRS), Munich,
Germany} and T. Gehren} \offprints{\\M. Bergemann, e-mail: masha@rejoin.net}
\institute{ Institute for Astronomy and Astrophysics, Ludwig-Maximilian
University, Scheinerstr. 1, 81679 Munich, Germany}
\date{Received date / Accepted date}
\abstract
{}
{Following our solar work, we perform NLTE calculations of the Mn abundance for
fourteen stars with [Fe/H] from 0 to -2.5, mainly to show how NLTE affects Mn
abundances in cool stars of different metallicities.}
{The spectrum synthesis and Mn abundances are based on statistical equilibrium
calculations using various estimates for the influence of hydrogen collisions.}
{The NLTE abundances of Mn in all studied stars are systematically higher than
the LTE abundances. At low metallicities, the NLTE abundance corrections may
run up to 0.5 - 0.7 dex. Instead of a strong depletion of Mn relative to Fe in
metal-poor stars as found by the other authors, we only find slightly subsolar 
values of [Mn/Fe] throughout the range of metallicities analyzed
here.}
{The [Mn/Fe] trend in metal-poor stars is inconsistent with the predictions
of galactic chemical evolution models, where Mn is less produced than Fe.}
\titlerunning{Mn in metal-poor stars}
\authorrunning{M. Bergemann \& T. Gehren}
\maketitle
\section{Introduction}
The abundances of manganese in stars of different populations have been
extensively studied during the past decades (Gratton 1989; Ryan et al.
\cite{RNB96}; Nissen et al. \cite{Nissen00}; Prochaska \& McWilliam
\cite{Proch00}; see also the review of Wheeler et al. \cite{Wheeler89}).
Although different methods and assumptions were applied to determine the Mn
abundances, they all agree that this element is deficient compared to iron in
metal-poor stars. The degree of the derived Mn depletion is claimed to vary from
the thin disk to the halo and bulge populations, and from cluster to field stars
(Gratton \cite{Gratton89}), and it is different for the Milky Way and the Sgr
dwarf galaxy (Prochaska \& McWilliam \cite{Proch00}).

Different assumptions are invoked to explain the trends, but they cannot
reconcile all observational facts simultaneously, especially when other elements
of the iron group (Cr, Co, and Ni) are considered. A critical issue is the main
astrophysical production site of manganese. While some favor the hypothesis of
a Mn overproduction with respect to Fe in supernovae of type Ia (Prochaska \&
McWilliam \cite{Proch00}; Nissen et al. \cite{Nissen00}, Sobeck et al.
\cite{Sobeck06}), McWilliam et al. (\cite{MR03}) suggest that
metallicity-dependent yields from type II supernovae can explain the observed
abundance trend. Both hypotheses imply that the Mn/Fe ratio in supernovae grows
with decreasing metal abundance.

No reliable constraint to Mn nucleosynthesis can be drawn from models of late
stellar evolution, because stellar yields depend on many factors that are poorly
known. As was shown among others by Umeda \& Nomoto (\cite{Umeda02}), the
position of the model mass cut influences the ratio of abundances originating
from the alpha-rich freeze-out and from the incomplete Si-burning. This affects
the abundances of Mn, Cr, and Co produced by the decay of $^{55}$Co, $^{52}$Fe,
and $^{59}$Cu, respectively. The explosion energy is another free parameter in
SN nucleosynthesis calculations (Heger \& Woosley \cite{Heger08}), which changes
the yields of a number of iron group elements simultaneously. Because of this,
the interpretation of yields becomes very non-trivial.

As noted above, different methods are used to calculate the Mn abundances. One
detail, however, is common to all analyses so far. The assumption of LTE, which
was shown to depart from reality for many elements, in particular for metal-poor
stars. Most of the NLTE investigations refer to the lighter elements, such as
C, N, O (Asplund et al. \cite{AS05},b; Takeda \& Honda \cite{TH05}) or Na, Mg,
and Al (Gehren et al. \cite{GS06}), but some also deal with heavier elements, 
among them Mashonkina et al. (\cite{LM07}: Ca), Korn et al. (\cite{KSG03}: Fe),
Bruls (\cite{Bruls93}: Ni), Mashonkina et al. (\cite{MGTB03}, \cite{MZG08}: Sr,
Ba and Eu). The purpose of this paper is to show the influence of NLTE effects
on the determination of Mn abundances for a sample of reference stars. At
present, we do not pretend to revise the observed trend completely, but even
from our small sample we conclude that the difference between LTE and NLTE Mn
abundances may become large in metal-poor stars. The use of LTE Mn abundances in
modeling Galactic chemical evolution is therefore not supported.

\begin{table*}
\begin{minipage}{\linewidth}
\renewcommand{\footnoterule}{}  
\tabcolsep1.7mm \small \caption{Stellar parameters and their estimated errors
for the selected sample. The final model NLTE and LTE abundances of Mn are given
in the last columns. Temperature and surface gravity errors are estimated,
whereas $\pi/\sigma_\pi$ represents the inverse fractional parallax error. $N$
is the number of Mn lines used for individual stars. See text for further
discussion of the data.}
\label{startab}
\begin{tabular}{lrr@{$\,\pm\,$}lr@{$\,\pm\,$}lcrrrlrr@{$\,\pm\,$}lr@{$\,\pm\,$}l
}
\hline\noalign{\smallskip}
Object\footnote{* marks the stars that were analyzed directly relative to the
Sun. The Mn abundances of all other stars are from analyses relative to
\object{HD 102200}, the results of which were added to the mean logarithmic 
abundance ratio [Mn/Fe] of this star relative to the Sun.} & HIP &
\multicolumn{2}{c}{$\Teff$} & \multicolumn{2}{c}{$\log g$} &
   $\xi_{\rm t}$ & [Fe/H] & [Mg/Fe] & $\pi/\sigma_\pi$ & Population & $N$ &
\multicolumn{4}{c}{[Mn/Fe]}\\
  &   & \multicolumn{2}{c}{[K]} & \multicolumn{2}{c}{ } & [km/s] &   &   &    &
  &   & \multicolumn{2}{c}{NLTE} &
    \multicolumn{2}{c}{LTE} \\
\noalign{\smallskip}\hline\noalign{\smallskip}
HD 19445         &  14594 & 5985 &  80 & 4.39 & 0.05 & 1.5 & $-1.96$ & 0.38 &
22.7 & Halo        & 10 & $-0.20$ & $0.03$ & $-0.52$ & $0.06$ \\
HD 25329         &  18915 & 4800 &  80 & 4.66 & 0.08 & 0.6 & $-1.84$ & 0.42 &
50.1 & Thick disk? &  6 & $-0.06$ & $0.02$ & $-0.23$ & $0.03$ \\
HD 29907         &  21609 & 5573 & 100 & 4.59 & 0.09 & 0.9 & $-1.60$ & 0.43 &
17.3 & Halo?       & 17 & $+0.05$ & $0.03$ & $-0.20$ & $0.03$ \\
HD 34328         &  24316 & 5955 &  70 & 4.47 & 0.07 & 1.3 & $-1.66$ & 0.42 &
14.4 & Halo        & 11 & $-0.10$ & $0.02$ & $-0.37$ & $0.03$ \\
HD 61421*        &  37279 & 6510 & 100 & 3.96 & 0.05 & 1.8 & $-0.03$ & 0.0  &
324.9 & Thin disk   & 17 & $-0.06$ & $0.04$ & $-0.08$ & $0.06$ \\
HD 84937         &  48152 & 6346 & 100 & 4.00 & 0.08 & 1.8 & $-2.16$ & 0.32 &
11.7 & Halo        &  7 & $-0.04$ & $0.02$ & $-0.42$ & $0.08$ \\
HD 102200*       &  57360 & 6120 &  90 & 4.17 & 0.09 & 1.4 & $-1.28$ & 0.34 &
10.5 & Halo        & 11 & $-0.04$ & $0.04$ & $-0.26$ & $0.05$ \\
HD 103095        &  57939 & 5110 & 100 & 4.69 & 0.10 & 1.0 & $-1.35$ & 0.26 &
140.0 & Halo        & 10 & $-0.07$ & $0.02$ & $-0.22$ & $0.03$ \\
BD$-4^\circ3208$ &  59109 & 6310 &  60 & 3.98 & 0.21 & 1.5 & $-2.23$ & 0.34 & 
3.7 & Halo        &  4 & $-0.12$ & $0.00$ & $-0.56$ & $0.09$ \\
HD 122196        &  68464 & 5957 &  80 & 3.84 & 0.11 & 1.7 & $-1.78$ & 0.24 & 
7.4 & Halo        & 12 & $-0.14$ & $0.03$ & $-0.47$ & $0.07$ \\
HD 122563        &  68594 & 4600 & 100 & 1.50 & 0.20 & 1.9 & $-2.51$ & 0.45 & 
5.2 & Halo        &  7 & $-0.15$ & $0.05$ & $-0.59$ & $0.10$ \\
HD 140283        &  76976 & 5773 &  60 & 3.66 & 0.05 & 1.5 & $-2.38$ & 0.43 &
18.0 & Halo        &  7 & $+0.02$ & $0.02$ & $-0.45$ & $0.10$ \\
G20-8            &  86443 & 6115 &  80 & 4.20 & 0.20 & 1.5 & $-2.19$ & 0.4  & 
5.1  & Halo        &  2 & $-0.02$ & $0.03$ & $-0.48$ & $0.03$ \\
HD 200580        & 103987 & 5940 &  80 & 3.96 & 0.06 & 1.4 & $-0.82$ & 0.46 &
13.8 & Thick disk  & 10 & $+0.26$ & $0.07$ & $+0.14$ & $0.07$ \\
\noalign{\smallskip}\hline\noalign{\smallskip}
\end{tabular}
\end{minipage}
\end{table*}
The paper is set as follows. The observed spectra and their reduction are
described in Sect. 2. The NLTE calculations and spectrum synthesis are
documented in Sect. 3. We present a considerable amount of details about the
methods of calculation, because this is very important for a judgement of the
resulting abundances. Methods of the abundance calculations and spectrum
synthesis are described in Sect. 4, and the main results are summarized in Sect.
5. They include the run of Mn/Fe abundance ratios with overall metallicity, a
comparison with some recent LTE results for Mn abundances, and their relation
to the chemical evolution of the Milky Way.
\section{Observations and stellar parameters}
Our sample consists of fourteen stars that were observed with the ESO UVES
echelle spectrograph at the VLT UT2 on the Paranal, Chile, in 2001, and with the
FOCES echelle spectrograph mounted on the 2.2m telescope of the CAHA observatory
on Calar Alto, during 1999 and 2000. For four stars (\object{HD 61421},
\object{HD 84937}, \object{HD 140283}, and \object{BD$-4^\circ3208$}), spectra
obtained with both telescopes were available. The UVES spectra have a
slit-determined resolution of $\lambda/\Delta\lambda \sim 50000$ and a
signal-to-noise ratio better than $S/N \sim 300$ near 5000 \AA. The data cover a
spectral range from 3300 to 6700 \AA, with a beam-splitter gap between 4470 and
4620 \AA. For the other stars, either only UVES spectra (\object{HD 29907},
\object{HD 34328}, \object{HD 102200}, \object{HD 122196}) or FOCES spectra
(\object{HD 19445}, \object{HD 25329}, \object{HD 103095}, \object{HD 200580})
were available. The latter have a resolution of $\sim 60000$ but an $S/N$ of
only $\sim 200$ near 5000 \AA, with the exception of \object{G20-8}, which was
observed with a resolution of only $\sim 40000$. The UVES observations of
\object{Procyon} (\object{HD 61421}) and \object{HD 84937} were taken from the
UVESPOP survey (Bagnulo et al. \cite{Bagnulo03}). All stars on our list were
observed at least twice. Table 1 lists the basic stellar parameters and the
abundance results for our stars. The individual population membership is taken
from Gehren et al. (\cite{GS06}), based on the [Al/Mg] ratio and stellar
kinematics. It is evident that most stars of the sample belong to the halo.

Parameters for most of the stars were taken from the analyses of Gehren et al.
(\cite{GLSZZ04}, \cite{GS06}), and Fuhrmann (\cite{KF04}). They were obtained
using the Balmer line profile fits of H$_\alpha$ and H$_\beta$ for the
determination of $\Teff$, a method quantified in Fuhrmann et al.
(\cite{Fuhrmann93}). The errors are obtained directly from profile fitting, and
they are estimated to stay within $\pm 50 \ldots 100$ K depending on a number of
assumptions that add to systematic errors (e.g., line broadening theory,
type of atmospheric models, treatment of convection). Generally, however, the
method is the most reliable among all of the methods available for the
temperature determination of solar-type stars. The calculation of the
surface gravity acceleration $\log g$ is based on {\sc Hipparcos} parallaxes and
on masses estimated from tracks of stellar evolution kindly provided by
VandenBerg et al. (\cite{DV00}) as indicated in Fuhrmann et al. (\cite{F97}) or
Gehren et al. (\cite{GLSZZ04}, \cite{GS06}). Using the common symbol [$X$] to
denote the logarithmic value of a stellar parameter $X$ relative to the
\object{Sun}, it follows that the errors propagate as $\Delta[g] =
\Delta[M]+2\Delta[\pi]+4\Delta[\Teff]$ with some minor additional error from
interstellar extinction and bolometric correction. Since the individual errors
in mass, parallax and temperature are largely independent, the errors of $\log
g$ for turnoff stars are caused almost entirely by the parallax errors,
whereas errors in mass determination or temperature are usually at least a
factor of $2$ smaller, and those from reddening or bolometric correction are
even less. Two exceptions to this rule are the cool main sequence stars
\object{HD 29907} and \object{HD 103095}, for which the \emph{mass}
determination is most uncertain. We note that the worst parallax of all stars in
our sample is that of \object{BD$-4^\circ3208$} with $\sigma_\pi/\pi = 0.27$;
this makes that particular surface gravity uncertain by more than $\pm 0.2$ dex.
The iron abundance and microturbulence velocities are derived from fitting
\ion{Fe}{ii} line profiles under the requirement that the Fe abundances are
independent of line strengths. The uncertainties in [Fe/H] and $\xi_{\rm t}$ are
estimated to be 0.05 dex and 0.2 \kms, respectively.

One and the same type of atmospheric model is used to determine stellar
parameters and perform spectrum synthesis. The models are line-blanketed LTE
atmospheres generated by the MAFAGS code (Fuhrmann et al. \cite{F97}). They use
the opacity distribution functions of Kurucz (\cite{Kurucz92}), and they 
reproduce the corresponding ATLAS9 models to within $< 20$ K. To account for the
high Fe abundance in these ODFs ($\log\varepsilon_{\rm Fe,\odot} = 7.67$), our
models are scaled down in overall metal abundance by $-0.16$ dex. Since with the
exception of \object{Procyon} our sample contains only thick disk and halo
stars, the ODFs for the rest of the sample were additionally scaled for an
$\alpha$-element overbundance of +0.4 dex, in rough agreement with previous 
analyses (Fuhrmann \cite{KF04}; Gehren at al. \cite{GS06}). A few of the stars 
are not easy to interpret. HD 25329 is nitrogen-rich, and HD 29907 is a 
single-lined binary. Both have an [Al/Mg] ratio that is typical of the thick 
disk rather than of the halo, but both have [Fe/H] abundances more typical of 
the Galactic halo. The asymmetrical drift velocity of HD 29907 deviates
significantly from the average thick-disk velocity distribution, and its
[Eu/Ba] ratio is also typical of the halo (Mashonkina et al. \cite{MGTB03}).
\section{Methods of NLTE calculations}
\begin{figure}
\resizebox{\columnwidth}{!}{\includegraphics{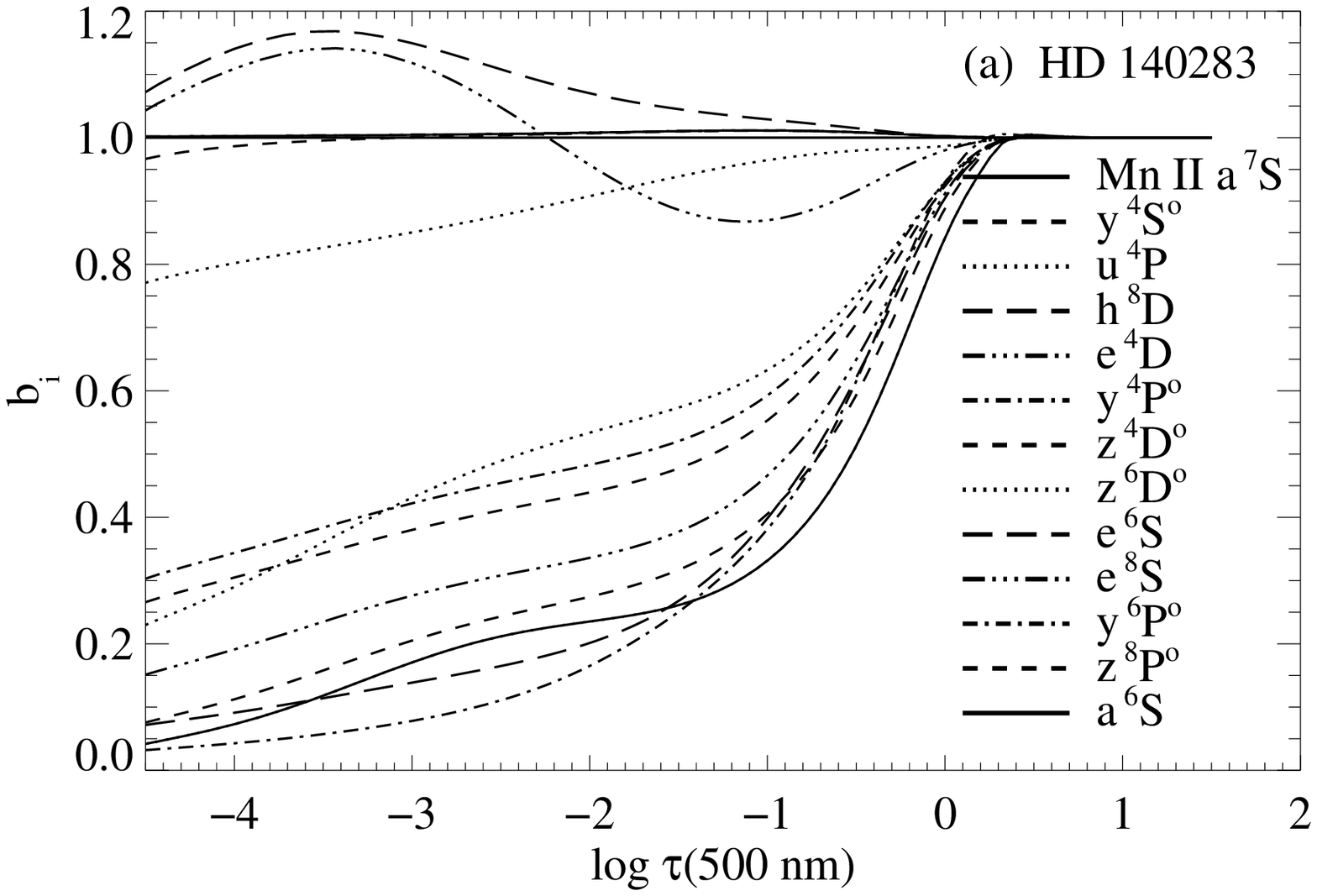}}\hfill
\resizebox{\columnwidth}{!}{\includegraphics{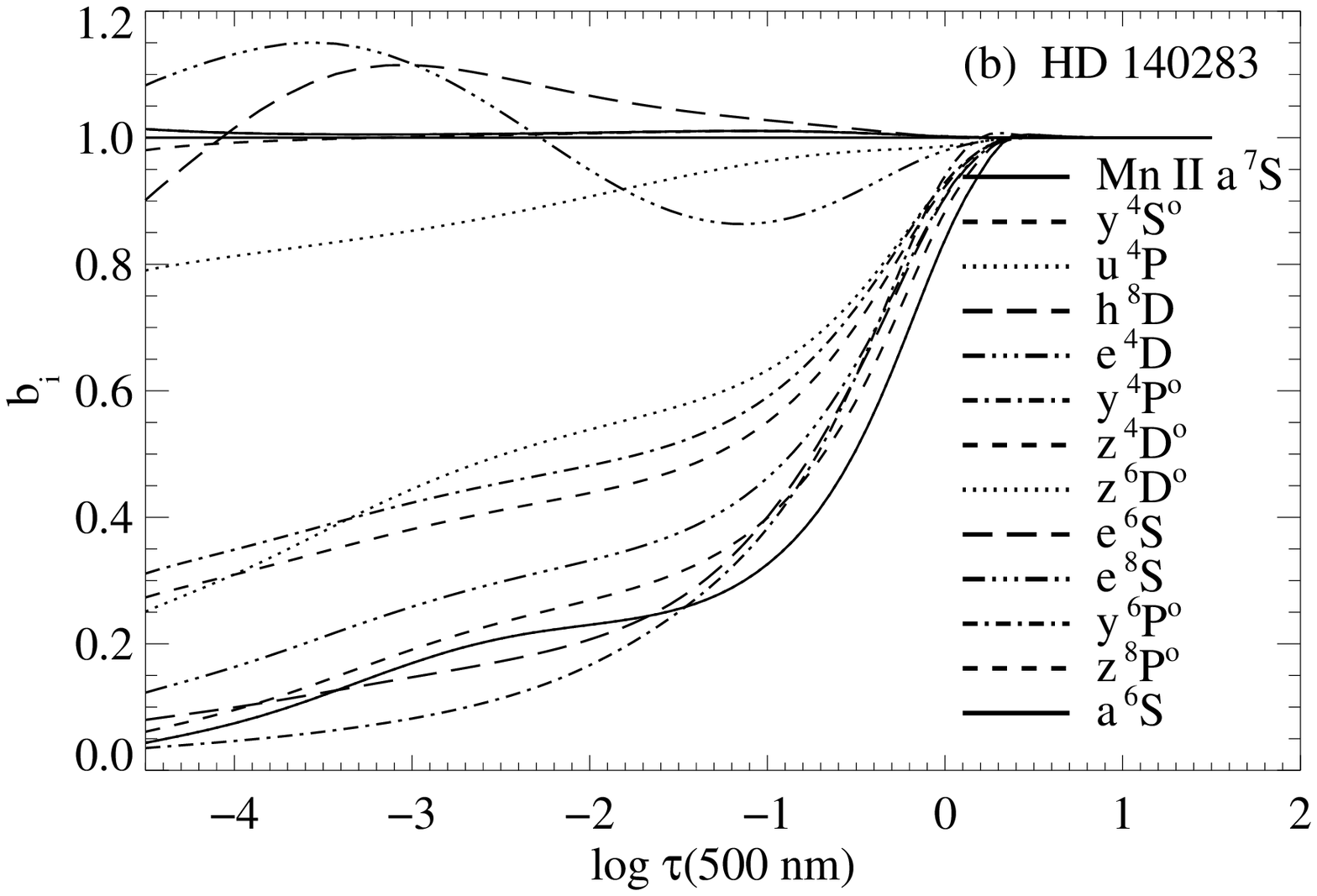}} \vspace{0mm}
\caption[]{Departure coefficients $b_i$ of selected \ion{Mn}{i} levels for
the metal-poor subgiant \object{HD 140283}: (a) reduced model with 245
levels of \ion{Mn}{i} and closed by the \ion{Mn}{ii} ground state; (b) full
model with 459 levels of \ion{Mn}{i}, \ion{Mn}{ii}, and \ion{Mn}{iii}.}
\label{hd140283}
\end{figure}

The NLTE level departure coefficients $b_i(\tau)$ are calculated using a Mn
model atom similar to that described in Paper I; minor modifications in the
model refer to the treatment of photoionization cross-sections and a number of
levels. The code DETAIL (Butler \& Giddings \cite{Butler85}) based on the method
of accelerated lambda iteration is used. We adopt the effective principal
quantum numbers in calculation of photoionization cross-sections. The model is
constructed with $245$ levels of \ion{Mn}{i} and closed by the ground state of
\ion{Mn}{ii}; it is by a factor of two smaller than the model atom investigated
in Paper I. Neglecting \ion{Mn}{ii} levels and \ion{Mn}{iii} ground state has
almost no influence on the behavior of \ion{Mn}{i} level departure coefficients.
Figure \ref{hd140283} shows that departure coefficients for the levels of
interest calculated for the most metal-poor star \object{HD 140283} are
identical over the whole optical depth scale. The difference in the models is
only reflected in the behavior of the \ion{Mn}{ii} ground state and
highly-excited \ion{Mn}{i} levels in the outer atmosphere. The main advantage is
that the time of computation is decreased significantly with the reduced atomic
model.
\begin{figure}
\vspace{-4mm}
\resizebox{\columnwidth}{!}{\includegraphics{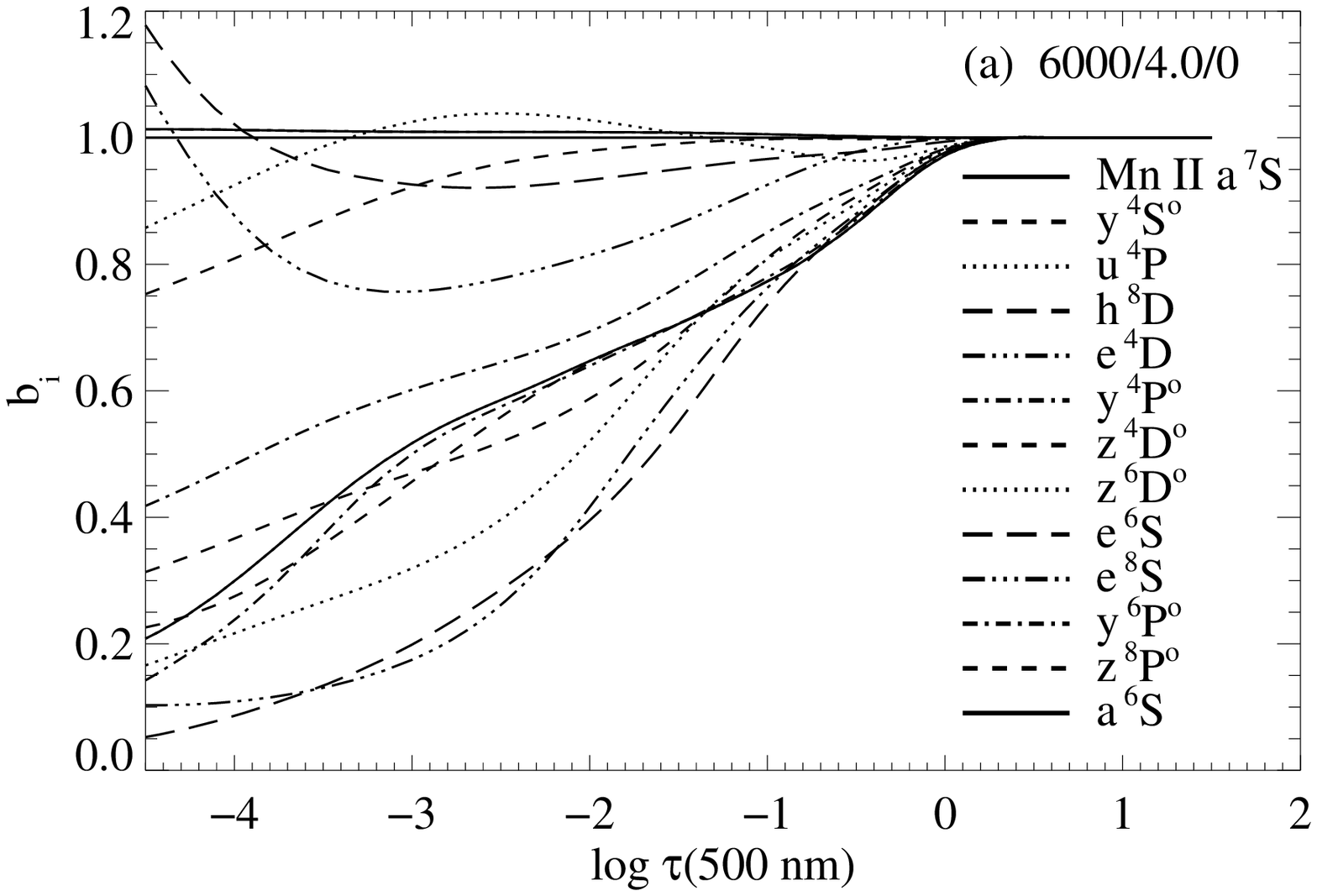}}\hfill
\vspace{-4mm}
\resizebox{\columnwidth}{!}{\includegraphics{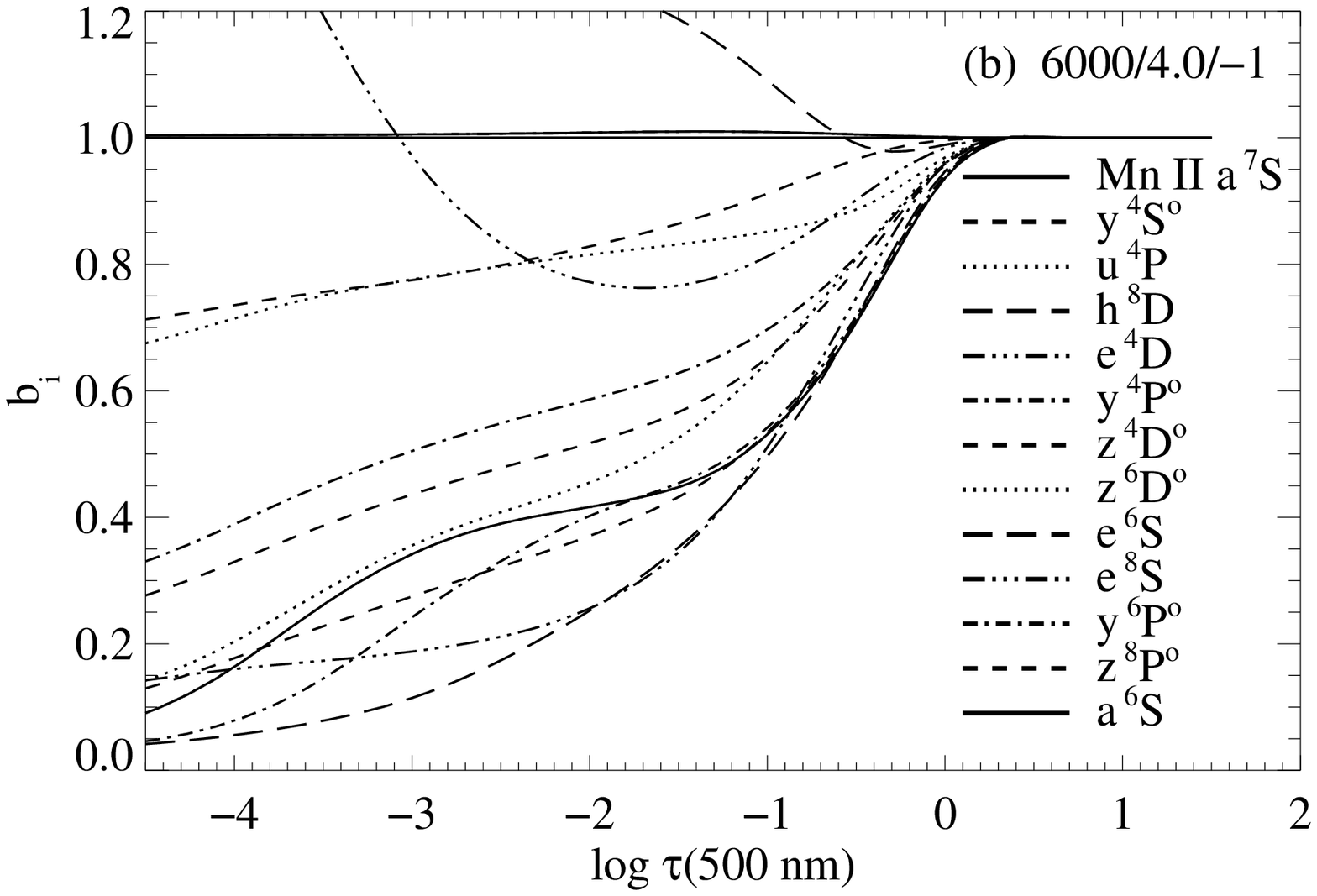}}\hfill
\vspace{-4mm}
\resizebox{\columnwidth}{!}{\includegraphics{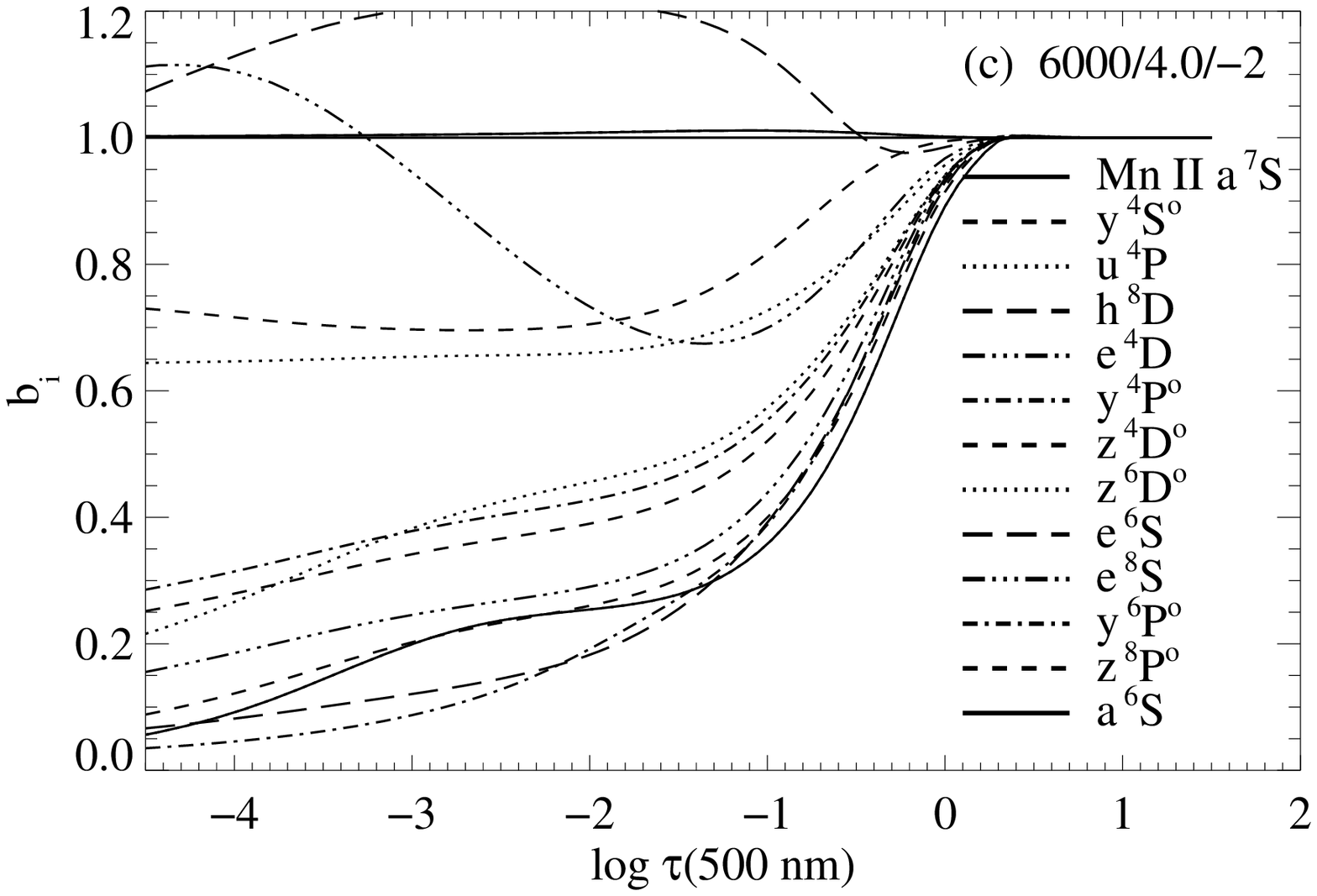}}\hfill
\vspace{-4mm} \resizebox{\columnwidth}{!}{\includegraphics{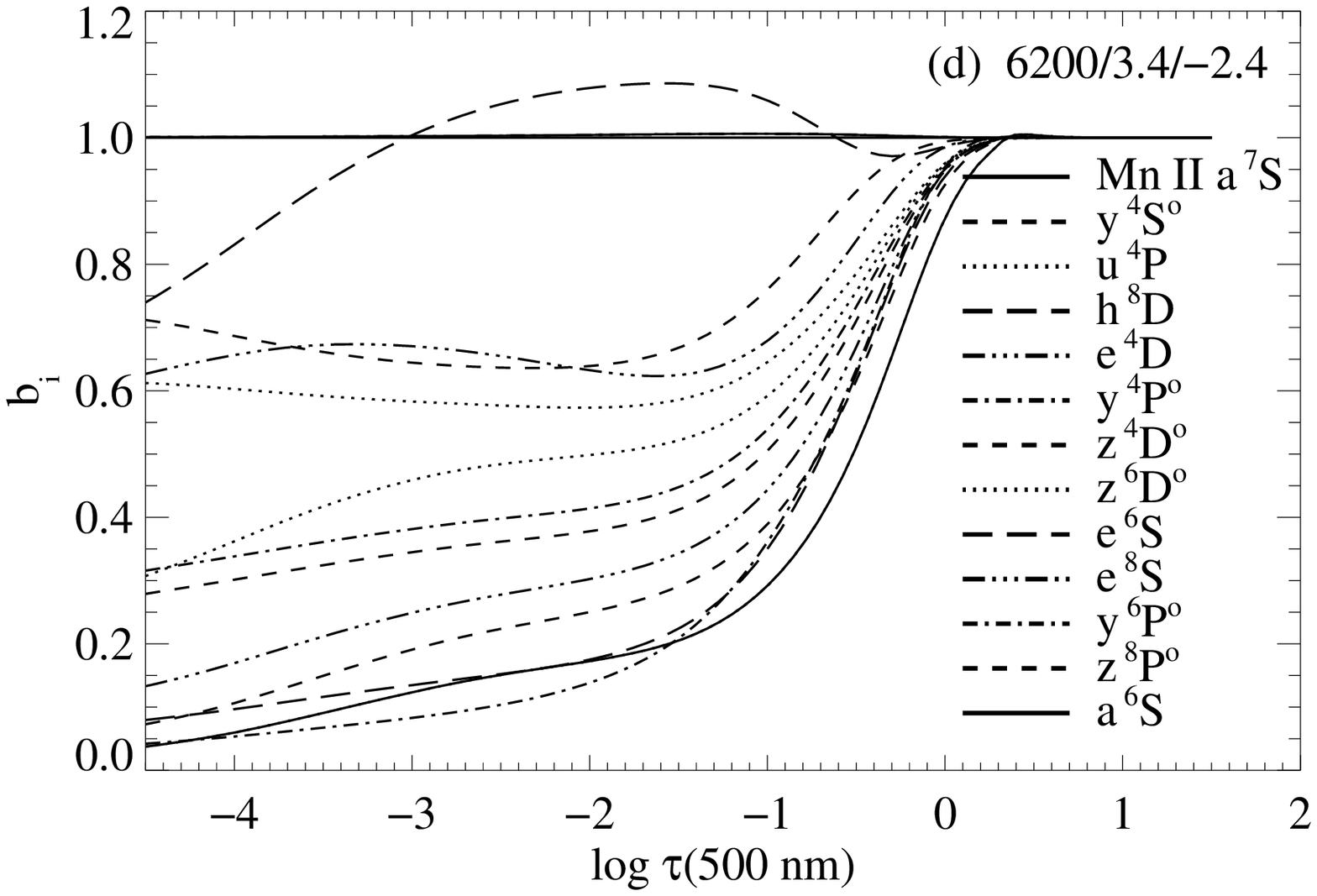}}
\vspace{-5mm} \caption[]{Departure coefficients $b_i$ of selected \ion{Mn}{i}
levels for stellar models with different parameters.} \label{reduced}
\end{figure}

We performed test calculations for selected models from a grid of theoretical
atmospheres with the purpose of investigating NLTE effects for \ion{Mn}{i} as a
function of stellar parameters. In this range, $T_\mathrm{eff}$ varies between
5000 and 6200 K, $3.0 \leq \log g \leq 4.6$, and $-3.0 \leq \mathrm{[Fe/H]} \leq
0.0$. The microturbulent velocity $\Vmic = 1 $ is assumed in all models. The
inelastic collisions with hydrogen atoms were neglected in the test calculations
($\SH = 0$). This results in slightly stronger NLTE effects compared to the 
case when hydrogen collisions are accounted for with our standard value ($\SH =
0.05$). On the other hand, a simple hydrogenic approximation for photoionization
underestimates the NLTE effects in \ion{Mn}{i} (see a discussion in Paper I). As
both assumptions work in opposite directions, we expect that our test models
are representative of the true NLTE effects acting in \ion{Mn}{i}. We show in 
Sect. 5 how variation in $\SH$ affects the abundances calculated from the
line-profile fitting.

The departure coefficients $b_i$ computed for four models of the grid with
different metallicities are shown in Figure \ref{reduced}. All \ion{Mn}{i}
levels involved in transitions of interest are underpopulated above $\log
\tau_{\rm 5000} \approx 0$. This is characteristic of all models irrespective of
the temperature, gravity, and metallicity. Hence the NLTE assumption generally
leads to an increase in the \ion{Mn}{i} line formation depths that forces a
weakening of the lines.

The NLTE effects become more pronounced with increasing effective temperature
and decreasing metallicity. In both cases, the UV radiation field is amplified
and increases the rates of b-b and b-f transitions in the \ion{Mn}{i} atom.
However, metallicity has a stronger influence on the populations of levels of
interest than effective temperature. The reason is that with increasing
temperature the collision rates also increase due to a larger kinetic energy of
the particles. This effect counteracts the NLTE influence of the radiation 
field. Whereas decreased metallicity implies a lower number density of metals,
which are the major suppliers of free electrons. As a result, the reduced
continuous opacity in the UV and decreased collision rates result in a strong
depopulation of all \ion{Mn}{i} levels.

In the solar metallicity model (Figure \ref{reduced}a), radiative transitions
with strong optical pumping effects are largely responsible for depopulation of
the low and intermediate energy levels. Photoionization from these levels is
inefficient, and that is the result of too low cross-section approximations
(see discussion in Paper I). In the metal-poor models (Figure \ref{reduced}b and
c), the dominant NLTE process is overionization from the low-excitation levels
\Mn{a}{6}{D}{}{}, \Mn{z}{8}{P}{\circ}{} and \Mn{a}{4}{D}{\circ}{} with
thresholds at $\lambda \sim 2330\ \AA$, $\lambda \sim 2406\ \AA$, and $\lambda
\sim 2727\ \AA$, respectively. Radiative b-b transition rates can no longer
compete with photoionization.

Variation of gravity has a marginal effect on the atomic level populations. For
the whole range of stellar parameters considered here, the change in gravity
from $\log g = 4.6$ to $\log g = 3.0$ results in a slightly weaker coupling of
the levels with each other. This is the result of a decreased collision
frequency, which tends to destroy LTE population ratios between the levels.
We show below that only at [Fe/H] $\leq -2$ and $\Teff \geq 6000$ K is the 
population of the \ion{Mn}{i} ground state is appreciably affected by the change
of gravity.

\subsection{NLTE and HFS effects on spectral lines and Mn abundances}
The abundance corrections $\Delta_\mathrm{NLTE}$ required to equalize the NLTE
and LTE \emph{equivalent widths} are given in Table \ref{grid}. These values
are calculated with the reference model of the Mn atom, but setting the
scaling factor for hydrogen collisions to 0. Table \ref{gridwithsh} shows the
NLTE abundance corrections for the same models calculated with $\SH = 0.05$.
Hyphens refer to the lines with theoretical NLTE equivalent widths below 3
$\mA$. We set this lower limit to the calculated $W_\lambda$ because our spectra
do not have a sufficient signal-to-noise ratio to measure the abundances from
such weak lines with an accuracy of less than 10 \%. For clarity, the 
discussion below is based on the data from Table \ref{grid} calculated with 
$\SH = 0$. Apart from the guesswork that collisions with neutral hydrogen atoms 
may become a main thermalizing mechanism in metal-poor stars, we know nothing 
about them.

The NLTE effects on a \ion{Mn}{i} line at $4033$ \AA\ for selected stellar
models from the grid are shown in Figure \ref{grid_figure}a. As in the previous
section, we select three models with different metallicities for comparison.
Other model parameteres are equal: $\Teff = 6000$ K and $\log g = 4$. 

The NLTE mechanisms, responsible for the behavior of line profiles at a solar 
metallicity, were explained in Paper I in detail. The strong lines are 
characterized by an amplified absorption in the core and a decreased absorption 
in the wings relative to LTE. The physics behind it is the overionization at
the depth of the line-wing formation, where $b_i < 1$ and $b_i < b_j$. At the 
depths of the core formation, photon losses in the line wings result in $b_i >
b_j$, so the core in NLTE is deeper than in LTE. The effect on the abundances is
relatively small. As can be seen in Table \ref{grid}, the NLTE abundance
corrections $\Delta_\mathrm{NLTE}$ vary from $+0.1$ to $-0.1$ dex for any model
with [Fe/H] $ = 0$.

With decreasing metallicity, line formation shifts to deeper layers, where
photoionization is dominant in depopulating all \ion{Mn}{i} levels. Maximum
$\Delta_\mathrm{NLTE}$ are found for the excited lines in the models with
$T_\mathrm{eff} \leq 5500$ K and [Fe/H] $\geq -2$.

With increasing $T_\mathrm{eff}$, the stellar flux maximum is shifted to the
shorter wavelengths, and ionization from the ground state of \ion{Mn}{i} becomes
more important. In the models with $T_\mathrm{eff} \geq 6000$ K and [Fe/H] $\leq
-2$, the NLTE abundance effect is more pronounced for the resonance triplet at
403 nm. However, as emphasized above, increasing $\Teff$ also increases the
rates of collisions, so maximum NLTE effects for the resonance 
lines are in fact found for the moderately warm model ($T_\mathrm{eff} \geq
5500$) with lowest metallicity (least amount of free electrons).

It is remarkable that, for all models with very low metal abundance except
$\Teff \sim 5000$ K, we find a difference of $\sim +0.2$ dex between the
NLTE corrections for resonance and excited lines. This difference is enhanced
for the model with lower gravity ($\log g = 3.4$). Thus, NLTE can solve the
discrepancy between the abundances derived from the \ion{Mn}{i} resonance
triplet at $403$ nm and excited lines, which is found in analyses of metal-poor
subdwarfs and subgiants (Gratton \cite{Gratton87}, Gratton \cite{Gratton89}, Lai
et al. \cite{Lai08}). The test calculations for the cool giant model ($\Teff =
4800$ K, $\log g = 1.8$, [Fe/H] $= -3.3$, not shown in Table \ref{grid}) also
indicate that the NLTE corrections for the resonance lines are by $\sim 0.15$
dex higher than those for excited lines at $4783$ and $4823$ \AA. This is the
first proof that the discrepancies between two line sets found by Johnson
(\cite{JJ02}) and Cayrel et al. (\cite{Cayrel04}) in their studies of giant
stars are due to NLTE.

\begin{figure}
\resizebox{\columnwidth}{!}{\rotatebox{-90}{\includegraphics{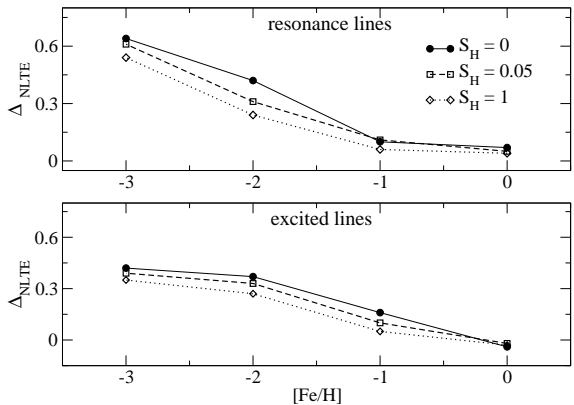}}}
\vspace{0mm} \caption[]{NLTE abundance corrections $\Delta_\mathrm{NLTE}$ (see
text) calculated with three different values of a scaling factor to inelastic
collisions with hydrogen $\SH = 0, 0.05, 1$: the average for the resonance
triplet at $403$ nm (top), the average for excited lines at $4783$ and $4823$
\AA. Calculations were performed for four models with $\Teff = 6000, \log g =
4,\mathrm{[Fe/H]} = 0,-1,-2, -3$.}
\label{nlte_abcor}
\end{figure}

It is useful to inspect the sensitivity of NLTE abundance corrections
$\Delta_\mathrm{NLTE}$ to the treatment of inelastic collisions with hydrogen.
This is a factor that can produce systematic errors in NLTE calculations. In the
following sections, we will show how the \textit{differential} [Mn/Fe] ratios
for the program stars react to changes in $\SH$. Here, we instead concentrate on
the change in NLTE abundance corrections derived for individual lines for
several models from the grid. Figure \ref{nlte_abcor} demonstrates the behaviour
of $\Delta_\mathrm{NLTE}$ calculated with three different scaling factors to the
standard Drawins formula $0, 0.05, 1$ (solid, dashed, and dotted lines,
respectively) as a function of a model metallicity ($\Teff = 6000$ K and $\log g
= 4$). Here, the average $\Delta_\mathrm{NLTE}$ derived for the resonance
triplet at $403$ nm are shown in the top panel, and that for the excited lines
at $4783$ and $4823$ in the bottom panel. These $five$ lines are traditionally
used in the analyses of Mn in metal-poor stars. The resonance lines are weakly
sensitive to H collisions at all metallicities, except for [Fe/H]$ \sim -2$. But
in the latter case, $\Delta_\mathrm{NLTE}$ is already large enough. The excited
lines steadily decrease in sensitivity to H collisions for decreasing Fe
abundance. This is expected because the major overionization in \ion{Mn}{i} is
expected to occur from the low-excited levels, e.g., $\Mn{z}{8}{P}{o}{}$, but
not from the ground state (see Paper I). In the metal-poor stars, overionization
is strongly amplified, as is obvious from the \textit{magnitude} of NLTE
corrections thus reducing the impact of collisions on level populations. The
important result following from Figure \ref{nlte_abcor} is that any collision
scaling factor chosen within the reasonable range (see below) leads to a
qualitatively similar behavior of NLTE abundance corrections. As a result, the 
error introduced by our poor knowledge of $\SH$ is systematic, much like the
errors in stellar parameters and models. With a \textit{free} choice of
$\SH$ from $0$ to $1$, one would most likely choose the value that produces the
NLTE abundance corrections lying in the middle of the upper and lower limits.
In this respect, $\SH = 0.05$ (our reference value) seems to be a good choice.
We show below that this value is also supported by the smallest fitted-abundance
spread between different lines of selected stars.

We have also checked the influence of hyperfine splitting (HFS) on \ion{Mn}{i}
line formation. Figure \ref{grid_figure}b shows NLTE profiles of \ion{Mn}{i}
line at $4823$ \AA\ calculated for three models with different metallicities
($\mathrm{[Fe/H]} = 0,-1,-2$, $\Teff = 5500$ K, $\log g = 4$). The synthetic
profiles computed with and without HFS are marked with solid and dotted lines,
respectively. The line is split into six HFS components. In the solar
metallicity model, the profile computed without HFS overestimates the abundance
by 0.35 dex. In the models with [Fe/H]$= -1$ and $-2$, the difference in
profiles corresponds to an abundance correction of 0.08 dex and 0.01 dex,
respectively. This example demonstrates the significance of HFS effects, which
must be treated correctly even when the abundance analysis of metal-poor stars
is performed.
\begin{figure}
\resizebox{\columnwidth}{!}{\includegraphics{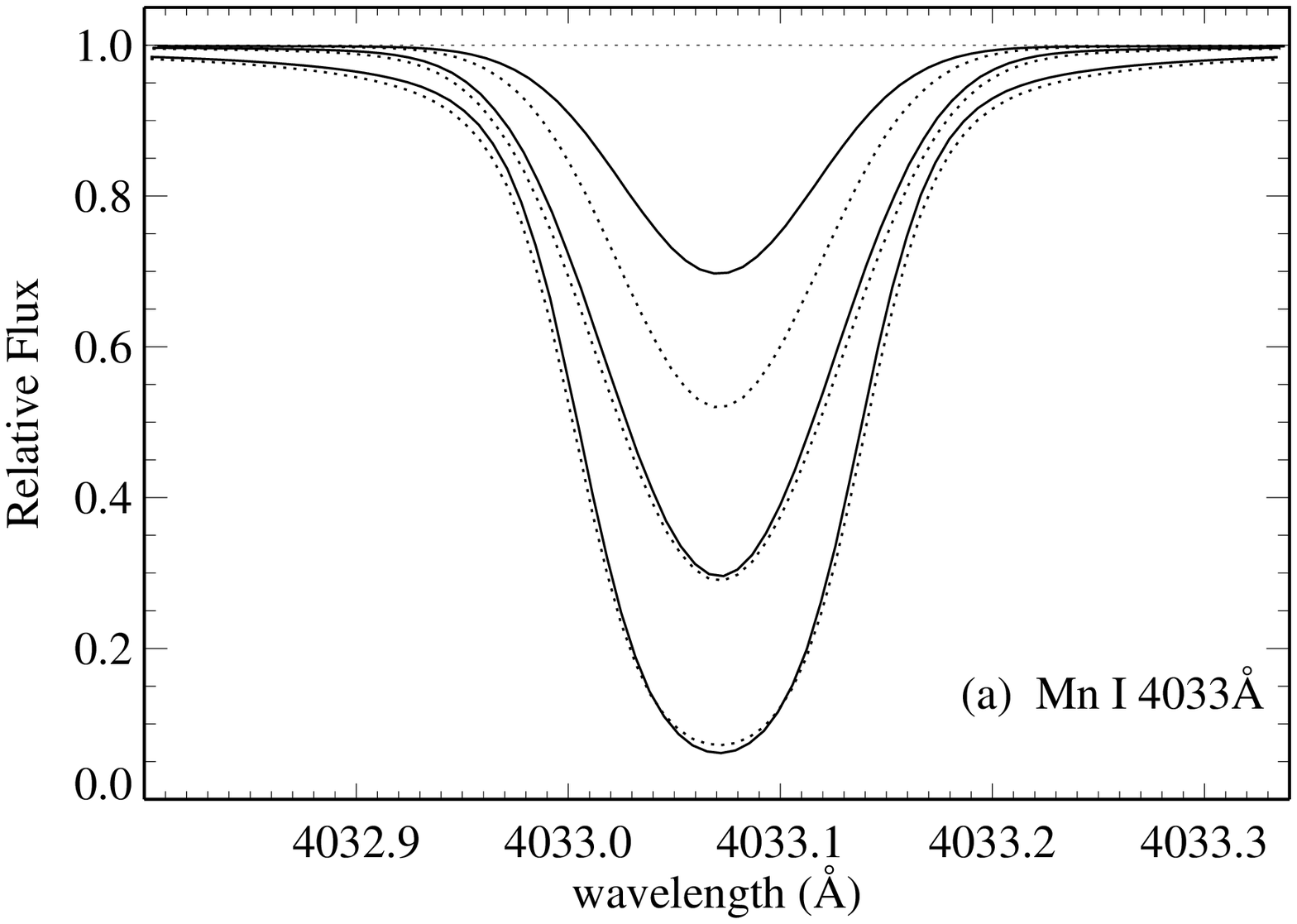}}\hfill
\resizebox{\columnwidth}{!}{\includegraphics{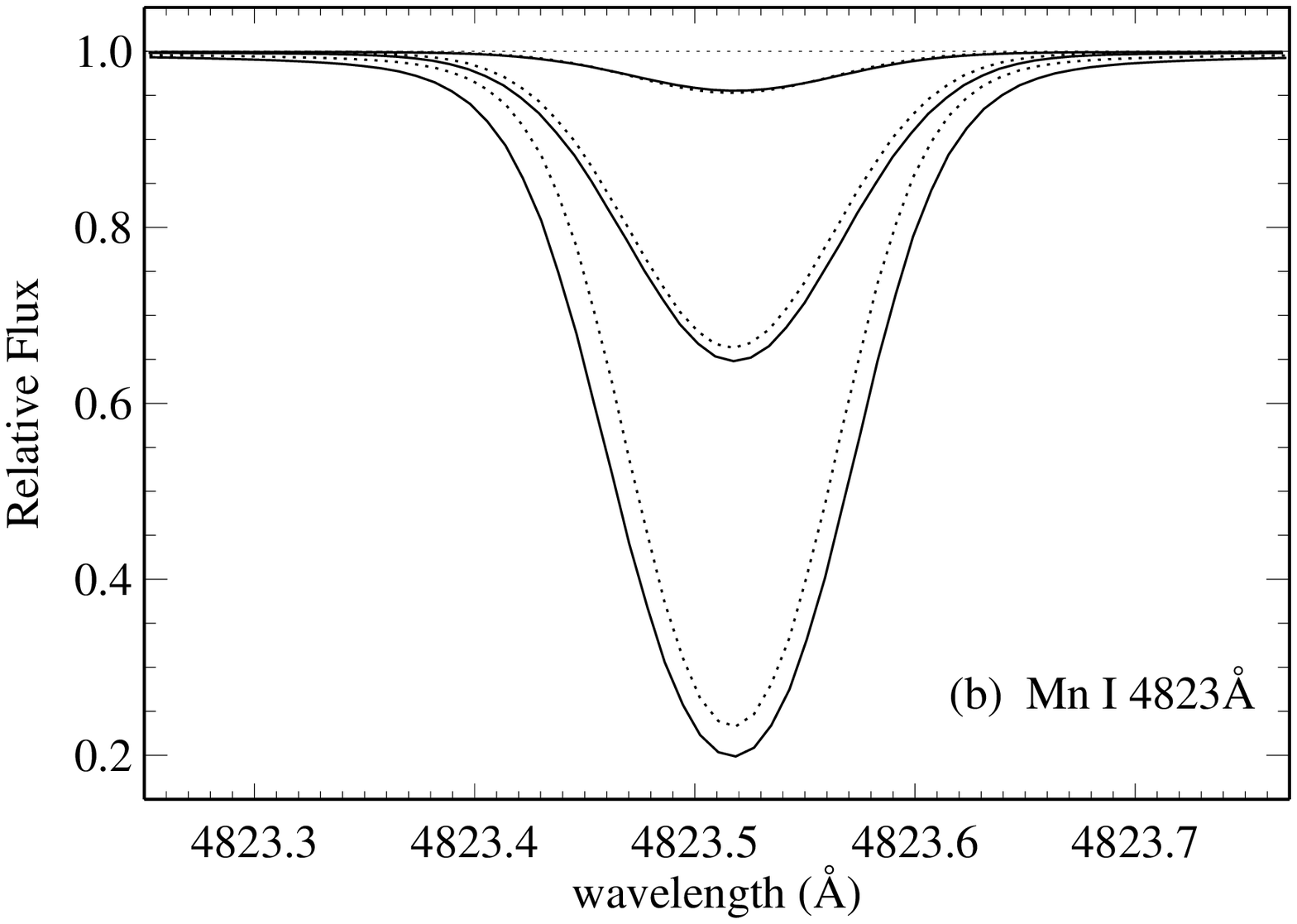}}
\vspace{0mm} \caption[]{Synthetic NLTE profiles of \ion{Mn}{i} lines for
different models from the grid. (a) NLTE effects on the line at 4033 $\AA$ for
the models with constant effective temperature and gravity ($\Teff =
6000, \log g = 4$) and different metallicities ($\mathrm{[Fe/H]} = 0,-1,-2$).
(b) HFS effects on the line at 4823 $\AA$ for the models with $\Teff = 5500,
\log g = 4,\mathrm{[Fe/H]} = 0,-1,-2$ (bottom to top).}
\label{grid_figure}
\end{figure}
\begin{table*}
\caption{ NLTE abundance corrections for selected models of the grid. 
Collisions with neutral hydrogen are neglected, $S_\mathrm{H} = 0$.}
\label{grid}
\tabcolsep1.2mm
\begin{tabular}{l|lllll|llll|lllll|llll}
\hline
$T_\mathrm{eff}$/$\log g$/ & \multicolumn{18}{c}{$\Delta_\mathrm{NLTE}$} \\
\cline{2-19}
\multicolumn{1}{c}{[Fe/H]} & \multicolumn{1}{c}{4018} & \multicolumn{1}{c}{4030}
&
\multicolumn{1}{c}{4033} & \multicolumn{1}{c}{4034} & \multicolumn{1}{c}{4041}
& \multicolumn{1}{c}{4055} & \multicolumn{1}{c}{4451} & \multicolumn{1}{c}{4739}
& \multicolumn{1}{c}{4754} & \multicolumn{1}{c}{4761} & \multicolumn{1}{c}{4762}
& \multicolumn{1}{c}{4765} & \multicolumn{1}{c}{4766} & \multicolumn{1}{c}{4783}
& \multicolumn{1}{c}{4823} & \multicolumn{1}{c}{6013} & \multicolumn{1}{c}{6016}
& \multicolumn{1}{c}{6021}
\\
\hline
5000/4/ 0 & 0.03 & 0.05 & 0.05 & 0.05 & 0.04 & 0.04 & 0.03 & 0.04 & 0.02 & 0.03
& 0.03 & 0.04 & 0.03 & 0.03 & 0.04 & -0.09 & -0.1 & -0.08 \\
5000/4/-1 & 0.15 & 0.15 & 0.15 & 0.15 & 0.14 & 0.15 & 0.2 & 0.24 & 0.09 & 0.25
& 0.18 & 0.24 & 0.24 & 0.13 & 0.12 & 0.17 & 0.16 & 0.13 \\
5000/4/-2 & 0.47 & 0.23 & 0.24 & 0.24 & 0.42 & 0.49 & 0.43 & - & 0.36 & - &
0.39 & 0.41 & 0.42 & 0.35 & 0.34 & - & - & - \\
5000/4/-3 & 0.54 & 0.45 & 0.45 & 0.49 & 0.56 & - & - & - & 0.48 & - & - & - & -
& 0.5 & 0.49 & - & - & - \\
5500/4/ 0 & 0.05 & 0.08 & 0.08 & 0.08 & 0.05 & 0.02 & 0.05 & 0.1 & 0.05 & 0.1 &
0.05 & 0.1 & 0.06 & -0.01 & -0.01 & 0.05 & 0.04 & 0.01 \\
5500/4/-1 & 0.28 & 0.1 & 0.15 & 0.15 & 0.26 & 0.25 & 0.26 & 0.27 & 0.15 & 0.29
& 0.22 & 0.3  & 0.29 & 0.14 & 0.14 & 0.24 & 0.24 & 0.22 \\
5500/4/-2 & 0.42 & 0.28 & 0.31 & 0.35 & 0.48 & 0.43 & 0.39 & - & 0.38 & - & 0.38
& 0.38 & 0.38 & 0.4 & 0.4 & - & - & - \\
5500/4/-3 & 0.51 & 0.69 & 0.7 & 0.72 & 0.51 & - & - & - & 0.5 & - & - & - & - &
0.52 & 0.52 & - & - & - \\
6000/4/ 0 & 0.1 & 0.07 & 0.07 & 0.08 & 0.05 &-0.02 & 0.05 & 0.11 & -0.02 & 0.11
& 0.05 & 0.1 & 0.08 & -0.03 & -0.04 & 0.06 & 0.05 & 0.01 \\
6000/4/-1 & 0.28 & 0.1 & 0.1 & 0.1 & 0.27 & 0.29 & 0.26 & 0.26 & 0.19 & 0.27 &
0.22 & 0.3  & 0.3  & 0.17 & 0.15 & 0.22 & 0.23 & 0.22 \\
6000/4/-2 & 0.38 & 0.4  & 0.42 & 0.42 & 0.38 & 0.41 & 0.36 & - & 0.35 & - &
0.34 & 0.35 & 0.35 & 0.37 & 0.37 & - & - & - \\
6000/4/-3 & - & 0.63 & 0.64 & 0.64 & 0.4 &- &- &- &- &- &- &- &- & 0.42 & 0.42
&- &- &- \\
6200/3.4/ 0   & 0.03 & 0.06 & 0.07 & 0.07 & 0.01 & 0.02 & 0.01 & 0.1 & -0.07 &
0.09 & 0 & 0.06 & 0.04 & -0.07 & -0.07 & 0 & -0.03 & -0.07 \\
6200/3.4/-1.2 & 0.32 & 0.1 & 0.1 & 0.15 & 0.31 & 0.38 & 0.32 & 0.28 & 0.27 &
0.28 & 0.28 & 0.31 & 0.31 & 0.25 & 0.25 & 0.25 & 0.24 & 0.25 \\
6200/3.4/-2.4 & 0.4  & 0.55 & 0.6  & 0.63 & 0.4  & 0.41 & 0.36 & - & 0.36 & - &
0.35 & 0.35 & 0.36 & 0.39 & 0.4 & - & - & - \\
6200/4.6/ 0   & 0.05 & 0.08 & 0.08 & 0.07 & 0.02 & 0.03 & 0.03 & 0.1 & -0.01 &
0.08 & 0.04 & 0.06 & 0.04 & 0.01 & 0.02 & 0.02 & 0 & -0.03 \\
6200/4.6/-1.2 & 0.28 & 0.15 & 0.15 & 0.15 & 0.26 & 0.31 & 0.28 & 0.26 & 0.24 &
0.26 & 0.25 & 0.28 & 0.28 & 0.23 & 0.22 & 0.23 & 0.23 & 0.23 \\
6200/4.6/-2.4 & 0.38 & 0.43 & 0.47 & 0.51 & 0.39 & 0.39 & 0.35 & - & 0.35 & - &
0.34 & 0.34 & 0.34 & 0.36 & 0.36 & - & - & - \\
\hline
\end{tabular}
\end{table*}
\begin{table*}
\caption{ NLTE abundance corrections for selected models of the grid. Collisions
with neutral hydrogen are included with a scaling factor $S_\mathrm{H} = 0.05$
(our reference value).}
\label{gridwithsh}
\tabcolsep1.2mm
\begin{tabular}{l|lllll|llll|lllll|llll}
\hline
$T_\mathrm{eff}$/$\log g$/ & \multicolumn{18}{c}{$\Delta_\mathrm{NLTE}$} \\
\cline{2-19}
\multicolumn{1}{c}{[Fe/H]} & \multicolumn{1}{c}{4018} & \multicolumn{1}{c}{4030}
&
\multicolumn{1}{c}{4033} & \multicolumn{1}{c}{4034} & \multicolumn{1}{c}{4041}
& \multicolumn{1}{c}{4055} & \multicolumn{1}{c}{4451} & \multicolumn{1}{c}{4739}
& \multicolumn{1}{c}{4754} & \multicolumn{1}{c}{4761} & \multicolumn{1}{c}{4762}
& \multicolumn{1}{c}{4765} & \multicolumn{1}{c}{4766} & \multicolumn{1}{c}{4783}
& \multicolumn{1}{c}{4823} & \multicolumn{1}{c}{6013} & \multicolumn{1}{c}{6016}
& \multicolumn{1}{c}{6021}
\\
\hline
5000/4/ 0 & 0.03 & 0.04 & 0.04 & 0.05 & 0.03 & 0.03 & 0.03 & 0.01 & 0.02 & 0.01
& 0.02 & 0.02 & 0.02 & 0.02 & 0.02 & -0.08 & -0.09 & -0.09 \\
5000/4/-1 & 0.15 & 0.1 & 0.1 & 0.1 & 0.13 & 0.11 & 0.15 & 0.19 & 0.07 & 0.2 
& 0.13 & 0.2  & 0.17 & 0.07 & 0.07 & 0.12 & 0.1  & 0.07 \\
5000/4/-2 & 0.41 & 0.23 & 0.23 & 0.23 & 0.36 & 0.43 & 0.37 & - & 0.31 & - &
0.34 & 0.37 & 0.37 & 0.3  & 0.3  & - & - & - \\
5000/4/-3 & 0.49 & 0.37 & 0.37 & 0.41 & 0.5  & - & - & - & 0.42 & - & - & - & -
& 0.45 & 0.44 & - & - & - \\
5500/4/ 0 & 0.02 & 0.05 & 0.04 & 0.04 & 0.02 & 0.03 & 0.01 & 0.04 & 0 & 0.03 &
0 & 0.03 & 0.01 & 0.01 & 0.01 & -0.08 & -0.12 & -0.1 \\
5500/4/-1 & 0.2 & 0.09 & 0.13 & 0.14 & 0.15 & 0.15 & 0.2 & 0.23 & 0.07 & 0.24
& 0.16 & 0.24 & 0.23 & 0.07 & 0.07 & 0.17 & 0.16 & 0.15 \\
5500/4/-2 & 0.41 & 0.27 & 0.28 & 0.29 & 0.42 & 0.39 & 0.42 & - & 0.33 & - &
0.33  & 0.34 & 0.34 & 0.35 & 0.35 & - & - & - \\
5500/4/-3 & 0.46 & 0.59 & 0.6 & 0.61 & 0.45 & - & - & - & 0.44 & - & - & - & - &
0.47 & 0.47 & - & - & - \\
6000/4/ 0 & 0.02 & 0.05 & 0.05 & 0.05 & 0.01 & 0.01 & 0.01 & 0.06 & -0.04 & 0.04
& 0.01 & 0.03 & 0.02 & -0.02 & -0.02 &-0.02 &-0.06 &-0.08 \\
6000/4/-1 & 0.21 & 0.11& 0.1 & 0.11& 0.18 & 0.21 & 0.2  & 0.2  & 0.13 & 0.21 &
0.17 & 0.22 & 0.22 & 0.1  & 0.1  & 0.16 & 0.16 & 0.17 \\
6000/4/-2 & 0.35 & 0.3 & 0.3 & 0.33 & 0.34 & 0.37 & 0.31 & - & 0.3  & - &
0.3  & 0.31 & 0.31 & 0.33 & 0.33 & - & - & - \\
6000/4/-3 & - & 0.6 & 0.61 & 0.61 & 0.38 &- &- &- &- &- &- &- &- & 0.38 & 0.39
&- &- &- \\
6200/3.4/ 0   & 0.01 & 0.05 & 0.06 & 0.07 &-0.03 & 0.01 &  0  & 0.08 & -0.1  &
0.07 & 0 & 0.07 & 0.03 & -0.08 & -0.1  & 0 & -0.03 & -0.07 \\
6200/3.4/-1.2 & 0.27 & 0.05 & 0.06& 0.11 & 0.26 & 0.32 & 0.26 & 0.24 & 0.22 &
0.24 & 0.22 & 0.25 & 0.26 & 0.2  & 0.2  & 0.2  & 0.19 & 0.2 \\
6200/3.4/-2.4 & 0.35 & 0.5  & 0.52 & 0.56 & 0.35 & 0.37 & 0.32 & - & 0.31 & - &
0.31 & 0.31 & 0.31 & 0.34 & 0.34 & - & - & - \\
6200/4.6/ 0   & 0.01 & 0.04 & 0.05 & 0.05 & 0.05 & 0.01 & 0.01 & 0.06 & -0.02 &
0.04 & 0.02 & 0.04 & 0.02 &-0.01 &-0.02 & 0 & -0.02 & -0.04 \\
6200/4.6/-1.2 & 0.22 & 0.1  & 0.1  & 0.11 & 0.2  & 0.25 & 0.22 & 0.21 & 0.18 &
0.21 & 0.19 & 0.22 & 0.22 & 0.17 & 0.16 & 0.17 & 0.17 & 0.17 \\
6200/4.6/-2.4 & 0.34 & 0.37 & 0.39 & 0.43 & 0.34 & 0.35 & 0.3  & - & 0.31 & - &
0.30 & 0.3  & 0.3  & 0.33 & 0.33 & - & - & - \\
\hline
\end{tabular}
\end{table*}
\section{Methods of abundance calculations}
\subsection{Spectrum synthesis}
Spectrum synthesis is carried out with the code SIU using the computed departure
coefficients $b_i(\tau)$ for \ion{Mn}{i} levels. For all other elements, LTE
line formation is assumed. The stellar line profiles are convolved with a
Gaussian of $\sim 3-4$ km/s. Rotational broadening and macroturbulence are
only treated separately for Procyon. Van der Waals damping constants $\log
C_{\rm 6}$ were determined in Paper I from solar line profile fitting. They are
on average $-0.1$ dex lower than the values calculated according to the Anstee 
\& O'Mara (\cite{Anstee95}) line broadening tables. Multiplet 22 with lines at
4451, 4453, and 4502 \AA\ was calculated with $\Delta \log C_{\rm 6} = -0.6$
relative to the Anstee \& O'Mara values. This correction was necessary to make
the Mn abundances derived from different lines of the multiplet agree with each
other.

Solar $\loggfesun$ values used here are slightly different from those published
in Bergemann \& Gehren (\cite{Bergemann07}) and Blackwell-Whitehead \& Bergemann
(\cite{Bl-Wh07})). This is due to an error in the treatment of hyperfine
splitting, which we found in our previous analyses. This error led to a
relatively large abundance scatter between different lines; however, the average
NLTE abundance of Mn remained the same, $\loge = 5.37 \pm 0.06$ dex.

The resonance triplet at 403 nm was calculated with $\log C_{\rm 6} = -31.2$
dex, that is, $0.65$ dex \emph{higher} than the values predicted by the Anstee
\& O'Mara theory. Also, lower values of microturbulent velocity are required to
fit these lines in the solar spectrum in contrast to all other lines, which are
well-fitted with $\Vmic = 0.9$ km/s. This is a result of huge line strengths;
e.g., the line at 4030 \AA\ has an equivalent width $W_\lambda \sim 325\ \mA$.
Such strong lines are formed over a large optical depth interval, and their 
profiles reflect the depth-dependent velocity distribution. The cores are formed
above $\log \tau_{\rm 5000} \sim -4.5$ being influenced by the
chromosphere. In this particular case, the NLTE theory produces cores that are
too narrow and deep compared to observations.

The line data used for spectrum synthesis are given in Table \ref{lines}.
Unless otherwise noted, the oscillator strengths are from Blackwell-Whitehead
\& Bergemann (\cite{Bl-Wh07}). The values of microturbulence velocities are 
listed in Table \ref{startab}.
\begin{table}
\renewcommand{\footnoterule}{}
\caption{Lines selected for abundance calculation. An asterisk in the $\log gf $
entry refers to the oscillator strengths taken from Booth et al.
(\cite{Booth84}). The $\loggfesun$ values for the five near-UV lines ($ 4018
\leq \lambda \leq 4041 \AA $) in the solar spectrum  cannot be reliably
calculated due to the line blending.}
\label{lines}
\begin{tabular}{lllllll}
\noalign{\smallskip}\hline\noalign{\smallskip}
~$\lambda$ & $\Elow$ & Lower & Upper & $\log gf $ & $\log C_{\rm 6}$ &
$\loggfesun$ \\
~[\AA] & [eV] & level & level &  &  & \\
\noalign{\smallskip}\hline\noalign{\smallskip}
4018.10 & 2.11  & \Mn{a}{6}{D}{ }{4.5} & \Mn{z}{6}{D}{o}{3.5} & --0.31*&
--31.0~ & - \\
4030.76 & 0.00  & \Mn{a}{6}{S}{ }{2.5} & \Mn{z}{6}{P}{o}{3.5} & --0.47*&
--31.2~ & - \\
4033.07 & 0.00  & \Mn{a}{6}{S}{ }{2.5} & \Mn{z}{6}{P}{o}{2.5} & --0.62*&
--31.2~ & - \\
4034.49 & 0.00  & \Mn{a}{6}{S}{ }{2.5} & \Mn{z}{6}{P}{o}{1.5} & --0.81*&
--31.2~ & - \\
4041.36 & 2.11  & \Mn{a}{6}{D}{ }{4.5} & \Mn{z}{6}{D}{o}{4.5} & ~~0.29*&
--31.0~ & - \\
4055.51 & 2.13  & \Mn{a}{6}{D}{ }{3.5} & \Mn{z}{6}{D}{o}{3.5} & --0.08 &
--31.0~ & 5.29~~ \\
4058.91 & 2.17  & \Mn{a}{6}{D}{ }{1.5} & \Mn{z}{6}{D}{o}{0.5} & --0.46 &
--31.0~ & 4.81~~ \\
4451.58 & 2.88  &  \Mn{a}{4}{D}{ }{3.5} & \Mn{z}{4}{D}{o}{3.5} & ~~0.13 &
--31.3~ & 5.59~~ \\
4453.00 & 2.93  &  \Mn{a}{4}{D}{ }{1.5} & \Mn{z}{4}{D}{o}{0.5} & --0.62 &
--31.3~ & 4.83~~ \\
4502.22 & 2.91  &  \Mn{a}{4}{D}{ }{2.5} & \Mn{z}{4}{D}{o}{3.5} & --0.43 &
--31.3~ & 4.96~~ \\
4709.71 &  2.88  &  \Mn{a}{4}{D}{ }{3.5} & \Mn{z}{4}{F}{o}{3.5} & --0.49 &
--30.74 & 5.36~~ \\
4739.09 &  2.93  &  \Mn{a}{4}{D}{ }{1.5} & \Mn{z}{4}{F}{o}{1.5} & --0.60 &
--30.71 & 5.34~~ \\
4754.02 &  2.27  &  \Mn{z}{8}{P}{o}{2.5} & \Mn{e}{8}{S}{ }{3.5} & --0.07 &
--30.7~ & 5.26~~ \\
4761.51 &  2.94  &  \Mn{a}{4}{D}{ }{0.5} & \Mn{z}{4}{F}{o}{1.5} & --0.27 &
--30.86 & 5.12~~ \\
4762.36 &  2.88  &  \Mn{a}{4}{D}{ }{3.5} & \Mn{z}{4}{F}{o}{4.5} & ~~0.30 &
--30.86 & 5.57~~ \\
4765.85 &  2.93  &  \Mn{a}{4}{D}{ }{1.5} & \Mn{z}{4}{F}{o}{2.5} & --0.08 &
--30.86 & 5.25~~ \\
4766.41 &  2.91  &  \Mn{a}{4}{D}{ }{2.5} & \Mn{z}{4}{F}{o}{3.5} & ~~0.11 &
--30.84 & 5.41~~ \\
4783.39 &  2.29  &  \Mn{z}{8}{P}{o}{3.5} & \Mn{e}{8}{S}{ }{3.5} & ~~0.06 &
--30.7~ & 5.41~~ \\
4823.46 &  2.31  &  \Mn{z}{8}{P}{o}{4.5} & \Mn{e}{8}{S}{ }{3.5} & ~~0.15 &
--30.7~ & 5.52~~ \\
6013.47 &  3.06  &  \Mn{z}{6}{P}{o}{1.5} & \Mn{e}{6}{S}{ }{2.5} & --0.43 &
--30.64 & 5.01~~ \\
6016.59 &  3.06  &  \Mn{z}{6}{P}{o}{2.5} & \Mn{e}{6}{S}{ }{2.5} & --0.25 &
--30.64 & 5.15~~ \\
6021.73 &  3.06  &  \Mn{z}{6}{P}{o}{3.5} & \Mn{e}{6}{S}{ }{2.5} & --0.12 &
--30.64 & 5.28~~ \\
\noalign{\smallskip}\hline\noalign{\smallskip}
\end{tabular}
\end{table}
\subsection{Some aspects of a differential stellar analysis}
As evident from our previous studies (Bergemann \& Gehren \cite{Bergemann07}),
the scatter among laboratory $gf$-values does not allow an accurate
determination of the solar Mn abundance. Hence, we perform a differential
analysis of stars with respect to the Sun\footnote{Every single abundance for a
line fit in a stellar spectrum is related to the solar abundance from the same
line}, which excludes the use of absolute oscillator strengths, but requires the
knowledge of the $\loggfe$ values for \emph{each line} under investigation. The
Mn abundance in a metal-poor star relative to the Sun is given by
$$ \mathrm{[Mn/H]}= \loggfestar - \loggfesun. $$

Five strong near-UV lines ($ 4018 \leq \lambda \leq 4041 \AA $) of \ion{Mn}{i}
cannot be used used for a reliable abundance analysis in the solar spectrum due
to a severe blending. Unfortunately, in our spectra of the metal-poor stars, 
only these lines and three lines of multiplet $16$ are detected. Hence, we 
define another ``reference'' star, \object{HD 102200}, with [Fe/H] $= -1.28$
dex, whose $\loggferef$ values are used for the differential comparison with the
more metal-poor stars. In this case, we calculate an average [Mn/H] from all
detected lines in spectrum of a metal-poor star relative to \object{HD 102200},
and then correct this value by an average [Mn/H] derived from the visual and
near-IR lines of \object{HD 102200} relative to the Sun.

Obviously, our analysis relies on the assumption that the \emph{relative} NLTE
abundances [Mn/H] deduced from near-UV lines of \object{HD 102200} do not
deviate from the abundances deduced from the other lines of this star. Usually,
the large scatter results from the incorrectly treated line blends, which have
a different dependence on the $T_{\rm eff}$ and $\log g$. Incorrect stellar
parameters, e.g., microturbulence and model structure, may decrease the accuracy
of the abundances. This problem can be partly solved by using the ratio of two
similar elements [Mn/Fe], since the same atmospheric models were used to obtain
iron abundances in the stars studied here (see description in Sect. 2).
\begin{figure}
\resizebox{\columnwidth}{!}{\includegraphics{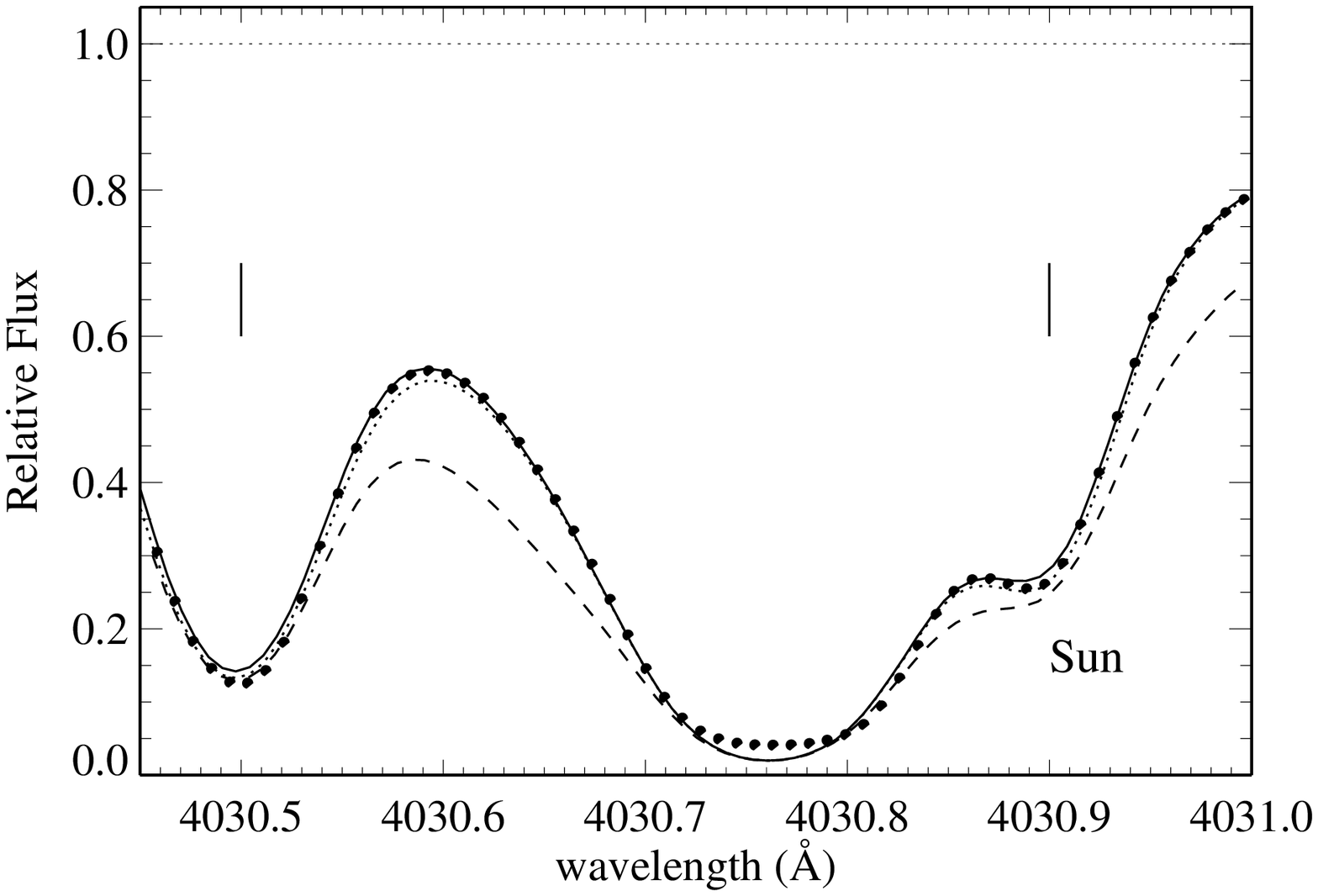}}
\resizebox{\columnwidth}{!}{\includegraphics{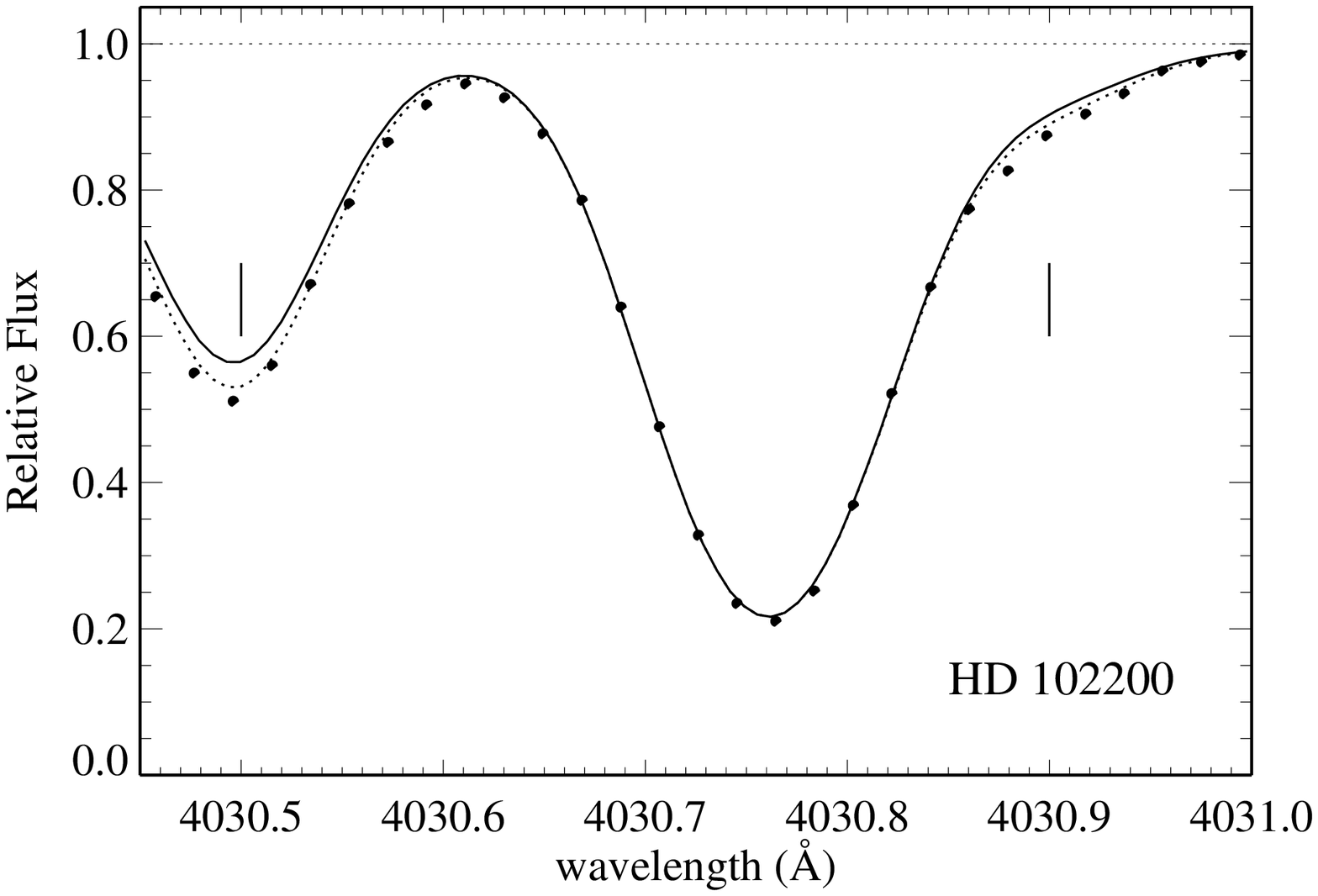}}
\resizebox{\columnwidth}{!}{\includegraphics{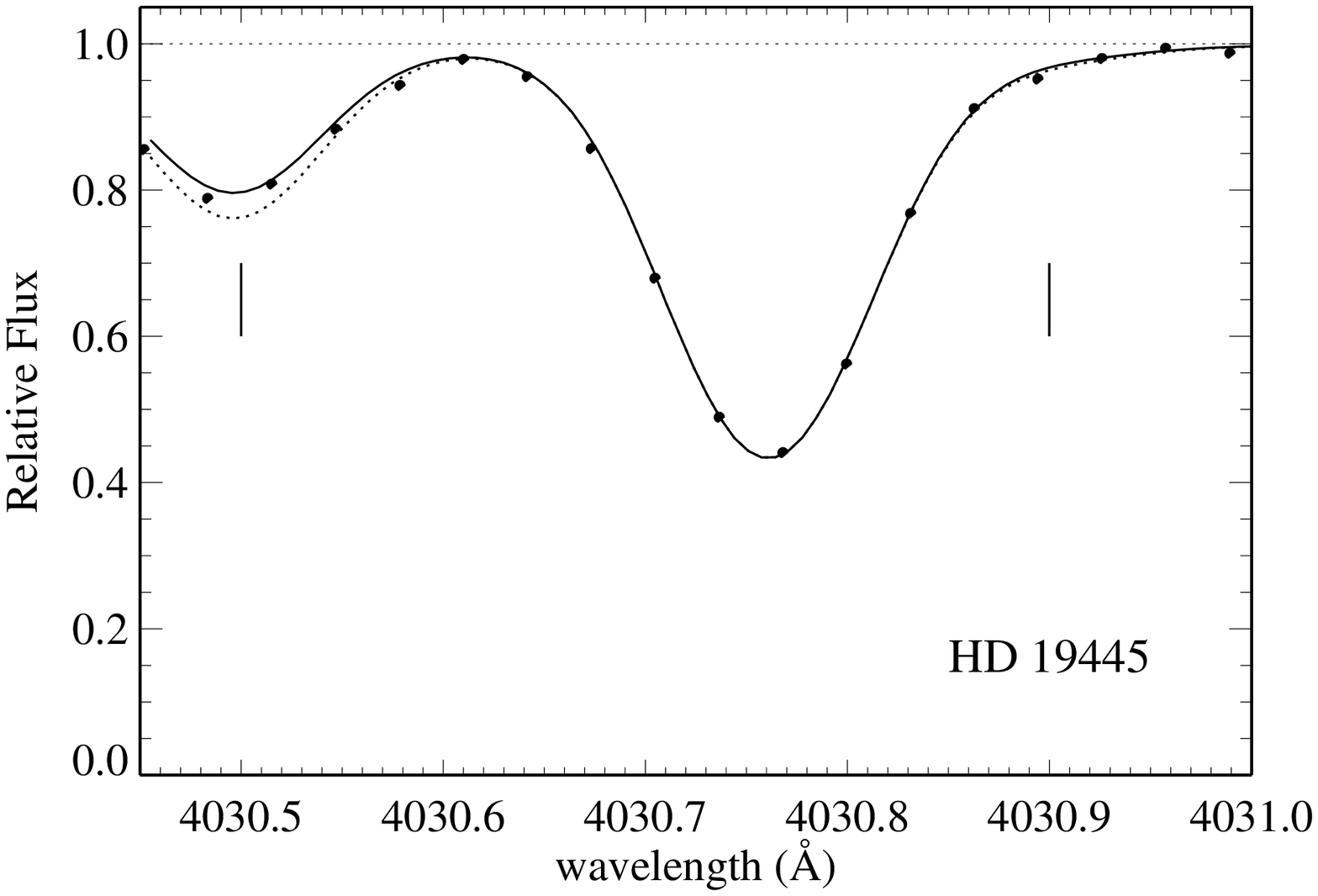}}
\vspace{0mm} \caption[]{The \ion{Fe}{i} blends in the blue and red wings of the
\ion{Mn}{i} line at 4030 \AA\ in spectra of the Sun, \object{HD 102200}, and
\object{HD 19445} (from top to bottom). The best theoretical NLTE profile
calculated with $\log C_{\rm 6} = -31.2$ is marked with a solid line, the same
profile with $\log gf$ for both \ion{Fe}{i} blends increased by 0.1 dex is shown
with a dotted line. The dashed line on the upper panel corresponds to the NLTE
profile calculated with $\log C_{\rm 6} = -30.5$.}
\label{blends}
\end{figure}

We have checked how blends contained in the near-UV \ion{Mn}{i} lines affect the
line profiles and Mn abundances derived from these lines in the Sun and in
metal-poor stars. Figure \ref{blends} shows the line at 4030 \AA\ in the spectra
of the Sun, \object{HD 102200}, and \object{HD 19445}, respectively. Vertical
lines mark the \ion{Fe}{i} blends in the red and blue wings. A solid line
represents the best theoretical NLTE profile, the dotted line shows the same
profile with $\log gf$ for both \ion{Fe}{i} blends increased by 0.1 dex. Even
this small correction produces a poor fit of the \ion{Mn}{i} line wings in the
solar spectrum; hence, we constrain the solar $\log gf$ values for the iron
blends at 4030.5 and 4030.9 $\AA$ to -0.8 and -1.6 dex, respectively. The
uncertainty of these values is $0.05$ dex. The profile of the same \ion{Mn}{i}
line in a metal-poor star is practically insensitive to a variation in iron line
strengths. We, thus, expect that the relative Mn abundances derived from the
near-UV lines in metal-poor stars are reliable.

It is noteworthy that the \emph{absolute} LTE and NLTE abundances from the
resonance triplet at 403 nm are lower than the abundances from the other lines
on average by $0.07 - 0.2 $ dex. This is a characteristic of all stars in our
sample. In the \emph{differential LTE analysis} relative to \object{HD 102200},
this inconsistency also appears and increases with decreasing metallicity. For
the metal-poor star \object{HD 140283}, we infer the $\Delta
\mathrm{[Mn/H]}_\mathrm{LTE}$ between the resonance and excited lines to be
$\sim 0.3$ dex. This finding supports the results of Cayrel et al.
(\cite{Cayrel04}). In their LTE analysis of Mn in metal-poor stars there is a
systematic difference of $\sim 0.4$ dex between the near-UV (resonance triplet
at 403 nm) and visual \ion{Mn}{i} lines (4754 and 4823 \AA).

The discrepancy is also present in a differential NLTE analysis, but even in the
most metal-poor stars it does not exceed 0.05 dex. We have good reason to
believe that the problem is concealed in the van der Waals damping constants;
namely, when the resonance lines are calculated with $\log C_{\rm 6} = -30.5$,
the abundance differences between resonance and excited lines completely vanish
for all stars in our sample. However, such a strong damping is cannot be
applied to fitting the solar resonance lines. The wings are too broad, and that
discrepancy cannot be removed by adjusting any other parameter in line synthesis
(see upper panel of Figure 4). Because of this, we decided to preserve the $\log
C_{\rm 6} = -31.2$ for the resonance lines, keeping in mind the possible error
of $0.05$ dex introduced by the uncertain $\log C_{\rm 6}$ value.

The problem elucidated above stems from the different sensitivities of strong
and weak lines to the van der Waals damping. A weak line in a metal-poor star
is strong in the solar spectrum, so its $\loggfe$ value will depend on the
adopted $C_{\rm 6}$ constant.

The most likely explanation for the discrepancy in \textit{absolute} line
abundances is that the oscillator strengths are erroneous. We use different sets
of experimental $gf$-values for the near-UV and visual lines, those of Booth et
al. (\cite{Booth84}) and of Blackwell-Whitehead \& Bergemann (\cite{Bl-Wh07}),
respectively. Cayrel uses the data of Booth et al. (\cite{Booth84}). The
comparison of Booth's values with the recent experiment of Blackwell-Whitehead
revealed significant disagreement for several strong \ion{Mn}{i} transitions. 

Finally, a differential analysis requires similar modeling of microscale 
velocities in the Sun and stars. In metal-poor stars, only a single 
depth-independent value of $\xi_{\rm t}$ can be recovered, so the constant 
microturbulence is adopted for the Sun, although its anisotropy and depth
dependence are now well-established (Gray \cite{Gray88}).
\section{Results}
The relative Mn abundance ratios [Mn/Fe] and their standard deviations $\sigma$ 
for all stars are listed in Table \ref{startab}. Stars with metallicities
higher than that of \object{HD 102200} were analyzed directly relative to the
Sun; in this case, the resonance triplet at 403 nm was not used for the reasons
given above. The Mn abundances in stars more metal-poor than \object{HD 102200}
are calculated relative to this reference star, and corrected by an average Mn
abundance derived for the \object{HD 102200} relative to the Sun. These
differential abundances are consistent within their $\sigma$ with Mn abundances
in stars calculated from the excited lines strictly relative to the Sun.

\begin{figure}[!ht] 
\resizebox{\columnwidth}{!}{\rotatebox{-90}{\includegraphics{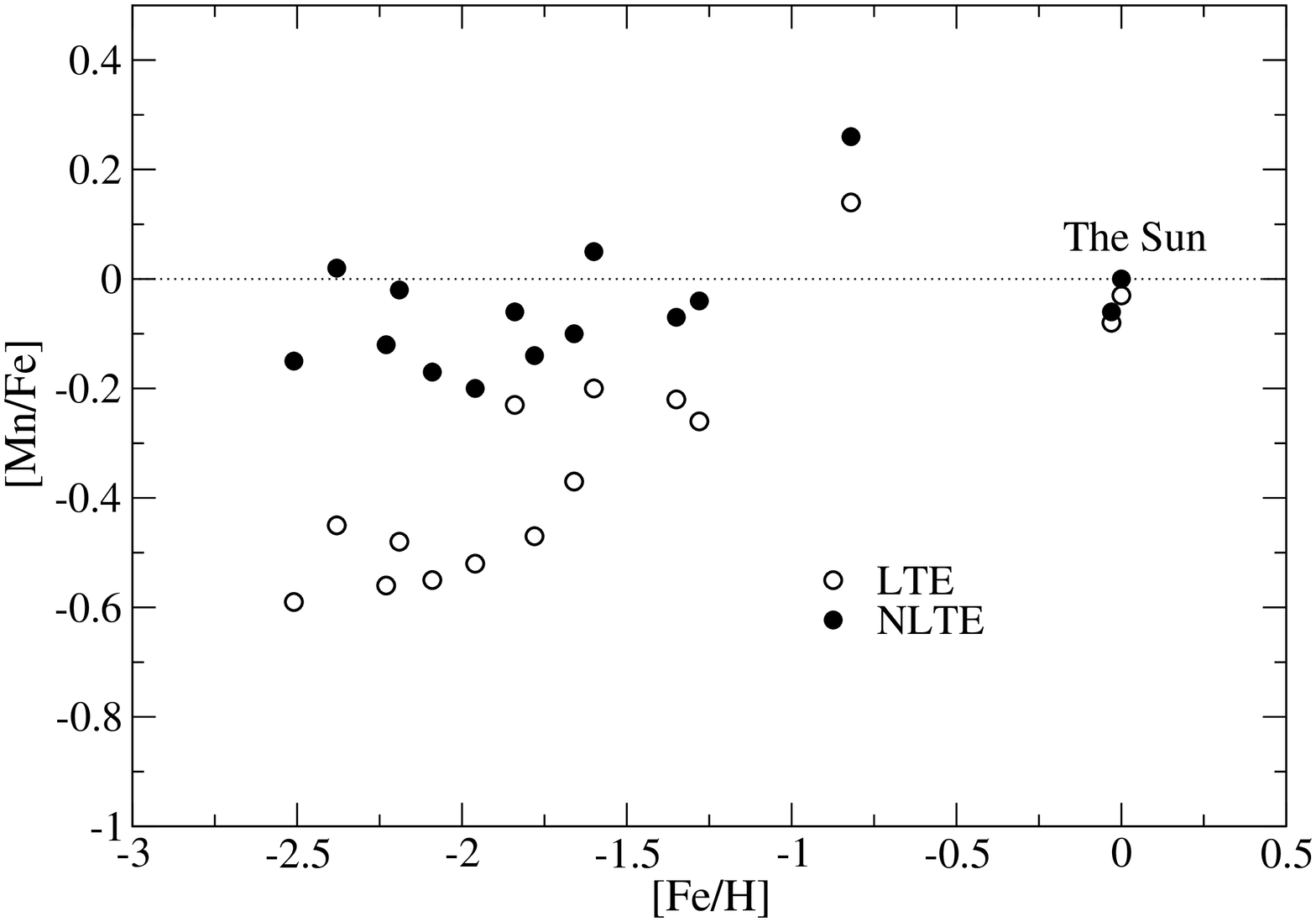}}}
\caption[]{[Mn/Fe] vs. [Fe/H] in metal-poor stars. NLTE abundances are marked
with filled circles. LTE abundances are marked with empty squares. The
spectroscopic binaries are shown with open circles.}
\label{abundall}
\end{figure}

Figure \ref{abundall} shows [Mn/Fe] ratios for all stars as a function of
metallicity. The binaries are marked with open circles. The most interesting
result is that the NLTE abundances of Mn in all stars are
\textit{systematically} higher than the LTE abundances; as expected, the
difference [Mn/Fe]$_\mathrm{NLTE}$ - [Mn/Fe]$_\mathrm{LTE}$ increases with
decreasing metallicity.

In the most metal-poor stars ([Fe/H] $\leq -2.5$), we find especially large NLTE
corrections for the resonance triplet at 403 nm. It is therefore to be expected
that LTE abundances of Mn in spectra of halo stars, where only the resonance
near-UV lines can be detected, are significantly underestimated. The dependency
of NLTE effects on (a) stellar parameteres ($T_\mathrm{eff}$, $\log g$, [Fe/H],
$\Vmic$) and (b) excitation potential of a line is obvious, and it is important
to keep it in mind even when LTE analysis is performed.

\begin{figure}
\resizebox{\columnwidth}{!}{\rotatebox{-90}{\includegraphics{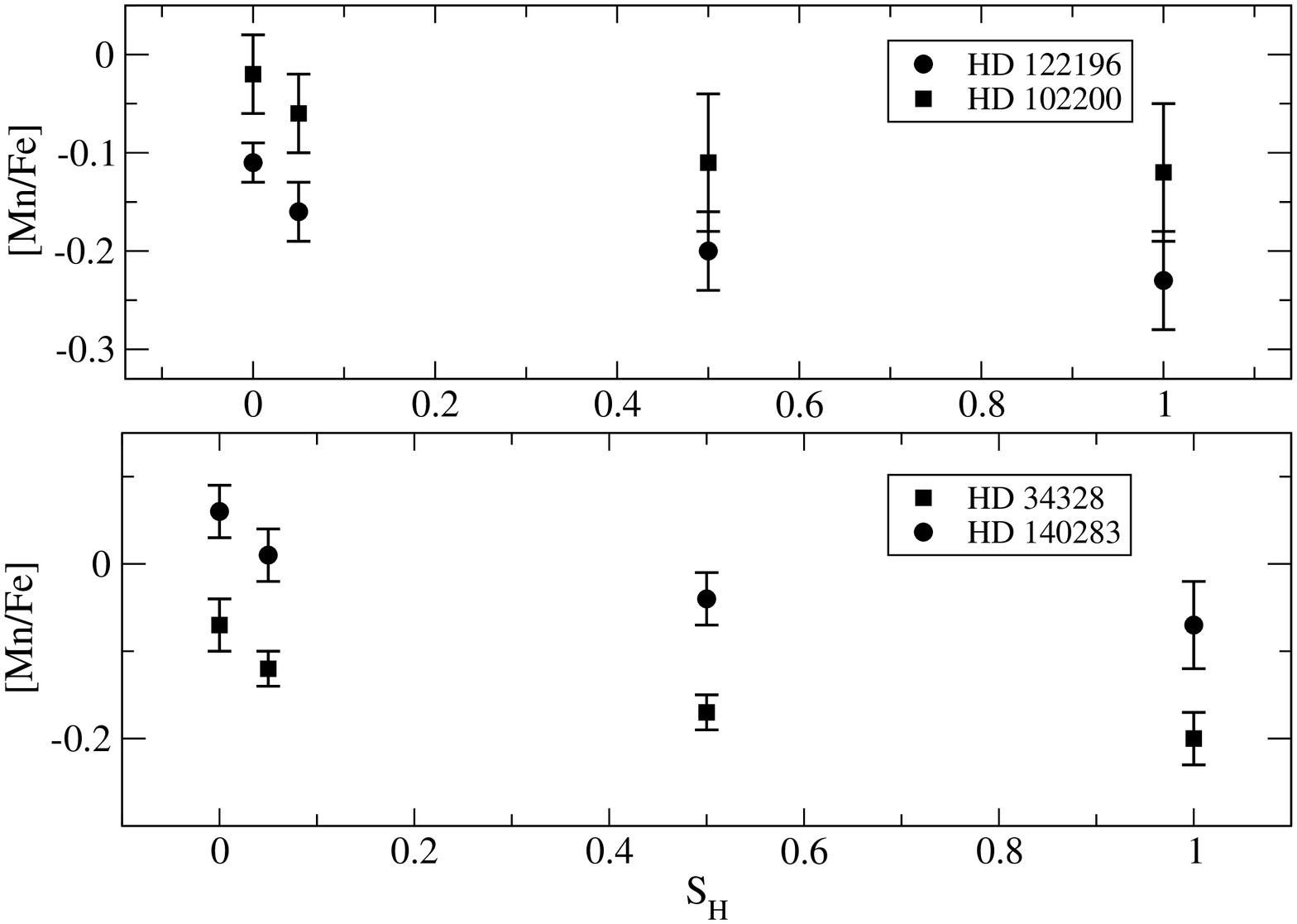}}}
\vspace{0mm} \caption[]{[Mn/Fe] ratios for \object{HD 34328}, \object{HD
102200}, \object{HD 140283}, and \object{HD 122196} calculated with $\SH = 0,
0.05, 0.5$, and $1$.} \label{sh34328}
\end{figure}

The most uncertain parameter in all NLTE calculations is $\SH$, the
scaling factor to the cross-sections for inelastic H collisions, which are
calculated with Drawin's (\cite{HD68}) formula. Fortunately, the [Mn/Fe] ratios
are not very sensitive to the choice of $\SH$, as long as very large scaling
factors ($\SH \gg 1$) are not used. The test calculations for four metal-poor
stars show (Figure \ref{sh34328}) that the change in a scaling factor $\SH =
0\ldots 1$ leads to a variation in the [Mn/Fe] ratios ($\Delta$[Mn/Fe]$_{0\ldots
1}$) in the range of $+0.05 \ldots -0.1$ dex. For stars with [Fe/H]$ > -1$, we
derive $\Delta$[Mn/Fe]$_{0\ldots 1} \sim +0.03 \ldots -0.05$. The error bars 
in Figure \ref{sh34328} denote standard  deviations $\sigma$ of [Mn/Fe] values.
Obviosuly, small $\sigma$ are achieved with $\SH \le 0.5$. It is reasonable to
suggest that $\SH = 0.05$ (our reference value) provides the smallest scatter
between different lines when all four stars are taken into account. This value 
agrees with the results derived for Ba and Eu by Mashonkina 
(\cite{Mashonkina96}) and for Mg emission lines by Sundquist et al.
(\cite{Sundqvist08}), who present convincing arguments in favor of very low 
($\ll 1$) but non-zero values for $\SH$. Also, according to Belyaev \& Barklem
(\cite{Belyaev03}), the standard Drawin's formula ($\SH = 1$) strongly
\textit{overestimates} the rate of transitions for Li. If these estimates are
true for Mn as well, we can claim only a slightly subsolar [Mn/Fe] trend with
decreasing metallicity. The variation in $\SH$ within the range $0\ldots 1$
will not change the general trend in Figure \ref{abundall} given the large
differences [Mn/Fe]$_\mathrm{NLTE}$ - [Mn/Fe]$_\mathrm{LTE}$ in metal-poor
stars.

We also expect that a thermalizing effect of H collisions in metal-poor
stars would be compensated by enhanced photoionization rates. In Paper I and
Paper II, we showed that a hydrogenic approximation for photoionization
cross-sections, which is the only one available to us, underestimates the NLTE
effects in \ion{Mn}{i}. The magnitude of underestimation increases in metal-poor
stars, where the radiation field is stronger than in the Sun. We performed test
NLTE calculations for the star \object{HD 34328} with photoionization
cross-sections for all levels increased by a factor of 300. The effect turned
out to be significant, because Mn abundances derived from the excited lines
increased by $0.4$ dex, on average. The abundances determined from the resonance
lines at $403$ nm only change by $+0.05$ dex. Strongly different NLTE
corrections required for the resonance and excited lines point at the inadequacy
of a simple hydrogenic photoionization approximation. Most likely,
cross-sections for particular low-excitation levels of \ion{Mn}{i} exhibit
resonance structures, as found for \ion{Fe}{i} (Bautista \cite{Baut97}).

\subsection{Comparison with recent work}

Recently, comparative manganese abundance analyses of metal-poor stars in
different populations have appeared in the literature, and \emph{all of them
refer to LTE}. The average Mn abundances for stars with metallicities $-1 \ldots
0$ are in agreement between the compiled analyses as to slope and absolute
values. We find a more or less smooth increase in [Mn/Fe] with [Fe/H] in the
results of Prochaska \& McWilliam (\cite{Proch00}), Nissen et al.
(\cite{Nissen00}), Reddy et al. (\cite{Reddy03}), Reddy et al. (\cite{Reddy06}),
and Feltzing et al. (\cite{FF07}). The Mn abundances in stars with lower
metallicities, [Fe/H] $< -1$, show a relatively large  scatter (Prochaska \&
McWilliam \cite{Proch00}; Bai et al. \cite{Bai04}; Cayrel et al.
\cite{Cayrel04}; Reddy et al. \cite{Reddy06}; Lai et al. \cite{Lai08}). But it
appears that the average [Mn/Fe] trend remains almost constant with the mean
value $-0.4$ dex.

Our NLTE Mn abundances are systematically larger than those from the comparison 
studies. The offset of $+0.2$ to $+0.4$ dex is mainly due to the NLTE effects, 
as can be predicted from Table \ref{gridwithsh}. In contrast, the Mn 
underabundance suggested by our LTE analysis is consistent with the results of 
the others. The LTE values of Prochaska \& McWilliam (\cite{Proch00}) corrected 
for mean NLTE effects from Table \ref{gridwithsh} agree reasonably well with
our NLTE Mn abundances in the metal-poor stars ($-2 <$ [Fe/H] $< -1$) at a 
constant level of [Mn/Fe] $\sim -0.1 \ldots 0.0$. Also, applying NLTE 
corrections to the data of Lai et al. (\cite{Lai08}), we can see that 
manganese follows the depletion of iron even in the most metal-poor stars. For
stars with $-2.5 <$ [Fe/H] $< -1.5$, our results slightly deviate from the
corrected LTE abundances presented by McWilliam et al. (\cite{MR03}). The reason
for this discrepancy is that in this metallicity range McWilliam et al.
(\cite{MR03}) used the LTE results of Johnson (\cite{JJ02}), which were obtained
by analyzing only \emph{giant} stars. Here, we note that work on incoherent data
compiled from various sources also tends to hide much of the chemical evolution
of the Galaxy in an exceedingly large scatter.

Unfortunately, the results presented by Feltzing et al. (\cite{FF07}) are for
disk stars alone, which are outside the range of metallicities covered by our
stellar sample. However, their finding of different abundance ratios [Mn/O] in
stars of thick and thin disks is not surprising, because it reproduces a similar
effect for the [Fe/Mg] ratio (Fuhrmann \cite{KF04}).

\subsection{Mn nucleosynthesis and chemical evolution}

Manganese is a \emph{neutron-rich} element in the sense that it has a
greater neutron-proton excess than the adjacent elements. It could therefore be
sensitive to the neutron excess available to the progenitor star in either
the carbon-burning or silicon-burning phases. Following the bimodal enrichment
scheme originally proposed by Arnett (\cite{WA71}; see also Woosley \& Weaver
\cite{WW82}), the yields of the neutron-rich elements always seem to depend on
the metals available to the star when the metal abundance of the progenitor
becomes larger than [Fe/H] $\sim -1.5$ (or [$\alpha$/H] $> -1.0$), as long as
the elements do not originate in a complete Si-burning environment. Therefore,
the stars with [Fe/H] $> -1.5$ must have been fed by SNe of a second or later
generation.

Taking into account possible residual systematic errors due to underestimating
photoionization by the hydrogen-like approximation, we can claim that [Mn/Fe] is
only slightly subsolar or even solar throughout the range of metal abundances
analyzed here. This would be confirmed by the results of Feltzing et al.
(\cite{FF07}) for the disk stars after applying the necessary NLTE corrections,
whereas it does not fit the earlier work of Gratton (\cite{Gratton89}) very
well. The models of chemical evolution presented by Timmes et al. (\cite{TWW95})
are, however, fully inadequate for representing the [Mn/Fe] ratio in metal-poor
stars. Based on the discussion in Umeda \& Nomoto (\cite{Umeda02}), it is
tempting to ascribe the discrepancy to the choice of the mass cut. Whether the
required increase in the mass cut or a change in the neutron excess is still
compatible with the observed Co abundances, will be examined in a future paper.
\begin {acknowledgements}
MB acknowledges with gratitude the Max-Planck Institute for Extraterrestrial
Physics (Germany) for her PhD fellowship.
\end {acknowledgements}

\end{document}